\documentclass[11pt]{article} 
\pdfoutput=1
\usepackage{hyperref}
\usepackage{amssymb}

\usepackage{amsmath}

\usepackage{amsthm}
\usepackage{dsfont}
\usepackage{libertine}
\usepackage[capitalise]{cleveref}
\usepackage{thm-restate}
\usepackage{subfig}
\usepackage{environ}
\usepackage{tikzsymbols}
\usepackage[normalem]{ulem}
\usepackage{multirow}
\usepackage{afterpage}

\makeatletter
\renewcommand{\section}{\@startsection {section}{1}{\z@}%
             {-3.5ex \@plus -1ex \@minus -.2ex}%
             {2.3ex \@plus .2ex}%
             {\normalfont\Large\scshape\bfseries}}
\renewcommand{\subsection}{\@startsection{subsection}{2}{\z@}%
             {-3.25ex\@plus -1ex \@minus -.2ex}%
             {1.5ex \@plus .2ex}%
             {\normalfont\large\scshape\bfseries}}
\renewcommand{\subsubsection}{\@startsection{subsubsection}{2}{\z@}%
             {-3.25ex\@plus -1ex \@minus -.2ex}%
             {1.5ex \@plus .2ex}%
             {\normalfont\normalsize\scshape\bfseries}}
\makeatother

\newcommand{\Description}[1] {}

\usepackage{mathtools}

\usepackage{array} 
\usepackage{booktabs}
\usepackage{tabularx,ragged2e}

\usepackage[ruled]{algorithm2e}
\usepackage{algpseudocode}
\usepackage{tikz}
\usepackage{mathrsfs}
\usepackage{enumerate}
\usepackage{bm}
\usetikzlibrary{arrows}
\usepackage{qcircuit}

\def\ve{\varepsilon}

\theoremstyle{plain}
\newtheorem{theorem}{Theorem}[section]
\newtheorem{lemma}[theorem]{Lemma}
\newtheorem{prop}[theorem]{Proposition}
\newtheorem{corollary}[theorem]{Corollary}

\theoremstyle{definition}
\newtheorem{definition}[theorem]{Definition}

\newtheorem{remark}[theorem]{Remark}
\newtheorem{fact}[theorem]{Fact}

\newcommand {\minusspace} {\: \! \!}

\newcommand {\fn} [2] {\ensuremath{ #1 \minusspace \br{ #2 } }}
\newcommand {\Fn} [2] {\ensuremath{ #1 \minusspace \Br{ #2 } }}

\newcommand {\set} [1] {\ensuremath{ \left\lbrace #1 \right\rbrace }}

\newcommand{\normthree}[1]{{\left\vert\kern-0.25ex\left\vert\kern-0.25ex\left\vert #1 \right\vert\kern-0.25ex\right\vert\kern-0.25ex\right\vert}}
\newcommand {\br} [1] {\ensuremath{ \left( #1 \right) }}
\newcommand {\Br} [1] {\ensuremath{ \left[ #1 \right] }}

\newcommand {\norm} [1] {\ensuremath{ \left\| #1 \right\| }}
\newcommand {\normsub} [2] {\ensuremath{ \norm{#1}_{#2} }}

\newcommand {\abs} [1] {\ensuremath{ \left| #1 \right| }}

\newcommand {\bra} [1] {\ensuremath{ \left\langle #1 \right| }}
\newcommand {\ket} [1] {\ensuremath{ \left| #1 \right\rangle }}
\newcommand {\ketbratwo} [2] {\ensuremath{ \left| #1 \middle\rangle \middle\langle #2 \right| }}
\newcommand {\ketbra} [1] {\ketbratwo{#1}{#1}}

\newcommand {\prob} [1] {\Fn{\Pr\,}{#1}}

\DeclareMathOperator*{\bigE}{\mathbb{E}}
\newcommand {\expec} [2] {\Fn{\bigE_{\substack{#1}}}{#2}}

\newcommand {\Tr} {\ensuremath{ \mathrm{Tr} }}
\newcommand {\partrace} [2] {\fn{\Tr_{#1}}{#2}}
\newcommand {\partracen} [2] {\fn{\tau_{\left[#1\right]}}{#2}}

\newcommand {\id} {\ensuremath{\mathds{1}}}

\usetikzlibrary{calc}
\tikzset{meter/.append style={draw, inner sep=10, rectangle, font=\vphantom{A}, minimum width=30, line width=.8,
 path picture={\draw[black] ([shift={(.1,.3)}]path picture bounding box.south west) to[bend left=50] ([shift={(-.1,.3)}]path picture bounding box.south east);\draw[black,-latex] ([shift={(0,.1)}]path picture bounding box.south) -- ([shift={(.3,-.1)}]path picture bounding box.north);}}}
 \usetikzlibrary{decorations.pathreplacing}

\NewEnviron{Anonymous}{\ifx\AnonymousSwitch\undefined\BODY\fi}

\graphicspath{ {images/} }

\newcommand{\Matrix} {\ensuremath{\mathcal{M}}}

\newcommand{\Holder} {H\"older}

\newcommand {\Base}[1] {\ensuremath{\mathcal{B}_{#1}}}
\newcommand {\degeps}[2] {\ensuremath{\widetilde{\operatorname{deg}}_{#1}\br{#2}}}
\newcommand {\QAC}[1][] {\ensuremath{\mathbf{QAC}^{#1}}}
\newcommand {\QACz} {\QAC[0]}

\newcommand {\QLCz} {\ensuremath{\mathbf{QLC}^0}}
\newcommand {\QNCz} {\ensuremath{\mathbf{QNC}^0}}
\newcommand {\AC}[1][] {\ensuremath{\mathbf{AC}^{#1}}}
\newcommand {\ACz} {\AC[0]}

\newcommand {\LCz} {\ensuremath{\mathbf{LC}^0}}

\newcommand {\ancillas} {\ensuremath{\psi}}

\newcommand {\polylog} {\ensuremath{\operatorname{polylog}}}

\newcommand {\opdilate}[1] {\ensuremath{#1^{\Uparrow}}}

\newcommand {\rx} {\ensuremath{\mathbf{x}}}
\newcommand{\Choi}[1] {\ensuremath{ \mathcal{J}(#1) } }
\newcommand{\Chois}[1] {\ensuremath{ \rho(#1) } }
\newcommand{\C} {\ensuremath{ \mathscr{C}} }
\newcommand{\supp}[1]{\ensuremath{\operatorname{supp}\br{#1}}}

\newcommand{\err} {\ensuremath{ \operatorname{err}} }
\newcommand{\EPR}[1] {\ensuremath{ \text{EPR}_{#1} } }

\newcommand {\CZGate} {CZ-gate}
\newcommand {\CZGr} {\ensuremath{\operatorname{CZ}}}
\newcommand {\CZG} {\ensuremath{\CZGr_n}}

\newcommand {\poly} {\ensuremath{\operatorname{poly}}}

\newcommand {\Parity}[1] {\ensuremath{\operatorname{Parity}_{#1}}}
\newcommand {\Parityn} {\Parity{n}}

\newcommand{\algname}[1]{\ensuremath{\textsc{#1}}}

\newcommand {\logO} {\ensuremath{\tilde{\mathcal{O}}} }
\usepackage[english]{babel}

\usepackage[letterpaper,top=2cm,bottom=2cm,left=3cm,right=3cm,marginparwidth=1.75cm]{geometry}

\usepackage{amsmath}
\usepackage{graphicx}

\title{Linear-Size \texorpdfstring{$\mathrm{QAC}^0$}{QAC0} Channels: Learning, Testing and Hardness}

\begin{Anonymous}
\author{
 Yangjing Dong\thanks{\scriptsize State Key Laboratory for Novel Software Technology, New Cornerstone Science Laboratory, Nanjing University, China. Email: dongmassimo@gmail.com.}
\and Fengning Ou\thanks{\scriptsize State Key Laboratory for Novel Software Technology, New Cornerstone Science Laboratory, Nanjing University, China. Email: reverymoon@gmail.com.}
\and Penghui Yao\thanks{\scriptsize State Key Laboratory for Novel Software Technology, New Cornerstone Science Laboratory, Nanjing University, China. Email: phyao1985@gmail.com.}~\thanks{\scriptsize Hefei National Laboratory, Hefei 230088, China.}
}
\end{Anonymous}

\begin{document}
\maketitle

\begin{abstract}
Shallow quantum circuits have attracted increasing attention in recent years,
due to the fact that current noisy quantum hardware can only perform faithful quantum computation for a short amount of time.
The constant-depth quantum circuits \QACz, a quantum counterpart of $\mathbf{AC}^0$ circuits, are the polynomial-size and constant-depth quantum circuits composed of only single-qubit unitaries and polynomial-size generalized Toffoli gates. The computational power of \QACz\ has been extensively investigated in recent years~\cite{10.5555/2011679.2011682, DBLP:journals/corr/abs-2005-12169, rosenthal:LIPIcs.ITCS.2021.32,NPVY24, ADOY24}.

In this paper, we are concerned with $\mathbf{QLC}^0$ circuits, which are linear-size \QACz\ circuits, a quantum counterpart of $\mathbf{LC}^0$. We provide a comprehensive study of $\mathbf{QLC}^0$ circuits. Our results are as follows.

\begin{itemize}
    \item We show that depth-$d$ \QACz\ circuits working on $n$ input qubits and $a$ ancilla qubits have
    approximate degree at most $\tilde{O}\br{(n+a)^{1-2^{-d}}}$,
    improving the $\tilde{O}\br{(n+a)^{1-3^{-d}}}$ degree upper bound of~\cite{ADOY24}.
    Consequently, this directly implies that to compute the parity function,
    \QACz\ circuits need at least $\tilde{O}\br{n^{1+2^{-d}}}$ circuit size.
    We obtain this bound by improving the techniques in~\cite{ADOY24} by using the techniques of unitary dilation and operator dilation.
       
    \item We present the first agnostic learning algorithm for \QLCz\ channels using subexponential running time and queries. Moreover, we also establish exponential lower bounds on the query complexity of learning \QACz\ channels under both the spectral norm distance of the Choi matrix and the diamond norm distance.

    \item We present a tolerant testing algorithm which determines whether an unknown quantum channel
    is a \QLCz\ channel.
    This tolerant testing algorithm is based on our agnostic learning algorithm.
    
\end{itemize}

Our approach leverages low-degree approximations of \QACz\ circuits and Pauli analysis as key technical tools. Collectively, these results advance our understanding of agnostic learning for shallow quantum circuits.
\end{abstract}

\newpage
\tableofcontents

\newpage

\section{Introduction}

Current quantum hardware suffers from decoherence and is only able to perform faithful quantum computation for a short amount of time.
This motivates the study of {\em shallow quantum circuits},
which have received increasing attention in recent years~\cite{bravyi2024classical,HLB+24,vasconcelos2024learning,arunachalam_et_al:LIPIcs.ICALP.2024.13}.
A simple yet important class of shallow quantum circuits are the \QNCz\ circuits,
which are polynomial size, constant depth quantum circuits with single-qubit and two-qubit quantum gates.
Despite these constrains, \QNCz\ circuits already demonstrate computational advantages over classical algorithms in specific tasks \cite{doi:10.1126/science.aar3106,10.1145/3313276.3316404,bravyi2020quantum,watts2023unconditional},
such as sampling from classically intractable distributions \cite{gao2017quantum,haferkamp2020closing,coble2022quasi}.
%Moreover, they are easier to implement on current noisy intermediate-scale quantum computers and well-suited for current experimental implementations.
However, due to the {\em light-cone} constraint, the ability of \QNCz\ circuits to
generate long-range entanglement or to compute Boolean functions is significantly limited. 

%Thus, researchers started investigating \QACz\ circuits, introduced by Green, Homer, Moore, and Pollett~\cite{moore1999quantum,moor} as a quantum counterpart of \ACz.
Green, Homer, Moore, and Pollett~\cite{moore1999quantum,moor} introduced \QACz\ circuits as a quantum counterpart of classical \ACz\ circuits, which are constant-depth polynomial-size quantum circuits consisting of single-qubit unitary gates together with arbitrarily size generalized Toffoli gates. Classical \ACz\ is a central object in circuit complexity, whose computational power is well understood and which delineates the frontier of current lower bound techniques. Thus, it is intriguing to investigate the computational power of \QACz. Recent progress on quantum hardware has also enabled the implementation of long-range multi-qubit operations, including the generalized Toffoli gate~\cite{rasmussen2020single,goel2021native,nikolaeva2025scalable} and the quantum fan-out gate~\cite{Gokhale2020QuantumFC}, underscoring the relevance of \QACz\ as a model for near-term quantum computation.

Using the Pauli analysis framework, Nadimpalli, Parham, Vasconcelos, and Yuen~\cite{NPVY24} 
characterized the structure of \QACz\ circuits of depth $d$ and ancilla size $n^{O(1/d)}$.
They showed that \QACz\ circuits
exhibit a low degree Pauli concentration similar to \ACz\ circuits when the ancilla is of size $n^{O(1/d)}$. 
Later, Anshu, Dong, Ou, and Yao~\cite{ADOY24} demonstrated that \QACz\ circuits with barely linear-size $n^{1+3^{-d}}$ have a small approximation degree.
Thus, linear-size \QACz\ circuits stand at the frontier of the lower bounds of quantum circuits. This motivates us to investigate the family of quantum circuits \QLCz\, as a quantum analog of \LCz~\cite{1663737},  which are linear size \QACz\ circuits.

\paragraph{Hardness}
A central challenge in studying \QACz\ circuits is to understand the computational power of \QACz.
In particular, given that \ACz\ circuits can not compute the parity function~\cite{10.1145/12130.12132},
a core problem is whether \QACz\ circuits are stronger than \ACz\ circuits and can compute the parity function.
This problem has been extensively studied in the past~\cite{10.5555/2011679.2011682, DBLP:journals/corr/abs-2005-12169, rosenthal:LIPIcs.ITCS.2021.32,NPVY24, ADOY24},
where they gave different ancilla lower bounds for \QACz\ circuits to compute the parity function.
Notably, given that the parity function has linear Fourier degree,
the low-degree approximation results of~\cite{NPVY24} and~\cite{ADOY24} show that \QACz\ circuits can not compute the parity function, if there is only $O\br{n^{1/d}}$ and $O\br{n^{1+3^{-d}}}$ ancilla, respectively.
Given that \QACz\ circuits with exponential ancilla can compute the parity function~\cite{rosenthal:LIPIcs.ITCS.2021.32},
a still open question is what is the minimal number of ancilla qubits required to compute the parity function.

An observation of~\cite{ADOY24} shows that any ancilla lower bound of the form $n^{1+\exp\br{-o(d)}}$ can be boosted to $n^c$ for any constant $c > 1$.
Thus if we can slightly improve the ancilla lower bound of~\cite{ADOY24},
then we can prove that any \QACz\ circuit can not compute the parity function.
So one possible path to resolve this open question is to improve the exponent in the $O\br{n^{1+3^{-d}}}$ lower bound.

\paragraph{PAC Learning and Agnostic Learning}
The task of learning an unknown quantum process, a.k.a. {\em Quantum process tomography},
is a fundamental task in quantum physics and quantum computation~\cite{Chuang_1997,Poyatos_1997,PhysRevLett.86.4195,O_Brien_2004,Scott_2008,PhysRevA.77.032322,Haah_2023}.
%A fundamental task in quantum physics and quantum computation is learning a quantum process, a.k.a. the quantum channel, a problem commonly referred to as {\em quantum process tomography}~\cite{Chuang_1997,Poyatos_1997,PhysRevLett.86.4195,O_Brien_2004,Scott_2008,PhysRevA.77.032322,Haah_2023}.
This problem has been extensively studied since the early days of quantum computation and has found broad applications in diverse areas, including quantum cryptography~\cite{Arapinis2021quantumphysical,doi:10.1126/science.adv8590}, quantum metrology~\cite{PhysRevLett.96.010401,Giovannetti2011}, and error mitigation~\cite{RevModPhys.95.045005}. 

Haah, Kothari, O’Donnell, and Tang~\cite{Haah_2023} have shown that learning an arbitrary quantum channel requires exponential resources. 
However, in practice we are not always faced with learning completely arbitrary quantum channels.
%Current quantum hardware suffers from decoherence and is only able to perform faithful quantum computation for a short amount of time, which can be modeled by {\em shallow quantum circuits}.
%Such circuits have received increasing attention in recent years~\cite{bravyi2024classical,HLB+24,vasconcelos2024learning,arunachalam_et_al:LIPIcs.ICALP.2024.13},
%due to their demonstrated computational advantages over classical algorithms in specific tasks \cite{doi:10.1126/science.aar3106,10.1145/3313276.3316404,bravyi2020quantum,watts2023unconditional},
%such as sampling from classically intractable distributions \cite{gao2017quantum,haferkamp2020closing,coble2022quasi}.
%Moreover, they are easier to implement on current noisy intermediate-scale quantum computers and well-suited for current experimental implementations.
Recently a number of learning algorithms for shallow quantum circuits have emerged in the literature~\cite{HLB+24,vasconcelos2024learning,arunachalam_et_al:LIPIcs.ICALP.2024.13,bao2024learning},
alongside many learning algorithms for quantum channels with other specific underlying structures,
such as junta channels~\cite{doi:10.1137/1.9781611977554.ch43,pmlr-v195-bao23b},
and Pauli channels~\cite{Harper_2020,Flammia_2020,Harper_2021,Chen_2022,Kunjummen_2023,wadhwa2024agnostic}.
%Instead, we may possess some prior knowledge about the channels in question.
%This has motivated  a line of research focused on learning quantum channels with , and others. A variety of algorithms have been proposed to efficiently learn such structured quantum channels.
In the standard model of PAC learning, it assumes that the target channels {\em exactly} satisfy the assumed structure. In real-world quantum systems, this assumption is often unrealistic due to unavoidable noise and imperfect gate operations. This motivates the study of {\em agnostic learning}, where the learning algorithm does not assume the target lies exactly within the hypothesis class.

Agnostic PAC learning, introduced by Kearns, Schapire, and Sellie~\cite{10.1145/130385.130424}, is a more general and robust learning model. It seeks to find a hypothesis within a given class that best approximates an arbitrary target function.  Agnostic learning has received significant attention since it was introduced. Despite its flexibility, designing agnostic learning algorithms with nontrivial performance guarantees remains a significant challenge, even when allowing sub-exponential running time~\cite{10.1145/3444815.3444825}.

The quantum version of agnostic learning for functions was first investigated by Arunachalam and de Wolf ~\cite{10.5555/3291125.3309633}, who established tight lower bounds on the sample complexity of agnostic learning in terms of the VC dimension. More recently, agnostic learning has been extended to quantum states, including product states~\cite{10.1145/3717823.3718207}, and stabilizer states \cite{chen2024stabilizerbootstrappingrecipeefficient,grewal2025agnostictomographystabilizerproduct}. Wadhwa, Lewis, Kashefi, and Doosti~\cite{wadhwa2024agnostic} further introduced {\em agnostic process tomography}, proposing algorithms for learning several classes of quantum channels, including bounded-gate circuits, Clifford circuits, Pauli strings, $k$-juntas, and low-degree circuits.
%In this work, we focus on agnostic learning for {\em shallow circuits}. 
%The agnostic tomography setting is hard in the sense that even for some simple quantum objects like product states and stabilizer states, it requires highly non-trivial algorithmic approaches 
%Furthermore, quantum process tomography -- a process crucial for verifying and characterizing quantum systems \cite{o2016efficient,10.1145/2897518.2897585}\pnote{correct references} -- faces inherent difficulties due to the no-cloning theorem, which prohibits perfect replication of unknown quantum states. \pnote{process tomography}
%Given their expressive power and experimental feasibility, a fundamental question arises:
%\begin{center}
%\textbf{How can we efficiently and agnostically learn a shallow quantum circuit ?}
%\end{center}

%To begin our discussion, we first revisit the classical counterpart of learning shallow circuits, a well-studied problem in computational learning theory.

%While the gate sets of \QACz\ and \ACz\ circuits differ, they share conceptual parallels: the Toffoli gate generalizes the OR gate and the single-qubit gates play a role similar to NOT gates. 

%Due to the incomparability of these results, we will present two different learning algorithms with incomparable parameters. The authors of \cite{HLB+24,vasconcelos2024learning} investigated further combinatorial properties of \QACz\ circuits and derived several related learning results based on these insights.

Despite this growing body of work, relatively little is known about agnostic learning for shallow quantum circuits. To the best of our knowledge, the only existing result prior to ours in this direction is due to Wadhawan, Lewis, Kashefi, and Doosti \cite{wadhwa2024agnostic} who initiated the study of agnostic learning in this setting.
We summarize all currently known results on learning shallow quantum circuits with many-qubit gates, including both upper and lower bounds on their agnostic and non-agnostic learnability in \cref{table: learning-on-shallow-circuits}. 

\begin{table}[tb]
\centering
\setlength{\extrarowheight}{4pt}
\begin{tabular}{|l|l|l|l|l|}
\hline
 & \textbf{Object } & \textbf{Model } & \textbf{Time } & \textbf{Sample } \\
\hline
%\cite{21923} & \begin{tabular}{l} \ACz\ Boolean function\end{tabular} & proper  & $O \br{n^{\log^d n}}$ & $O \br{n^{\log^d n}}$  \\
%\hline
%\cite{doi:10.1137/1.9781611975482.42} & \begin{tabular}{l} \LCz\ Boolean function \end{tabular}& proper agnostic & $2^{\logO \br{n^{1-2^{-d}}}}$ & $2^{\logO \br{n^{1-2^{-d}}}}$  \\
%\hline
\cite{NPVY24} & \begin{tabular}{l} \QACz\ $n \to m$ Channel,\\$a \leq O(\log n)$, Frobenius norm\footnotemark\end{tabular} & --  & $2^n$ & $\mathcal{O}  \br{n^{\log^d n}}$ \\
\hline
\cite{wadhwa2024agnostic} & \begin{tabular}{l} \QACz\ $n \to 1$ Channel,\\$a \leq \mathcal{O} (\log n)$, Frobenius norm\end{tabular} &  Agnostic & $\mathcal{O} \br{n^{\log^d n}}$ & $\mathcal{O}  \br{n^{\log^d n}}$  \\
\hline
%\cite{HLB+24} & \begin{tabular}{l} \QNCz\ Unitary,\\ diamond norm\end{tabular} & proper  & $\text{poly}(n)$ & $O(n^2 \log n)$  \\
%\hline
%\cite{HLB+24} & \begin{tabular}{l} \QNCz\ State,\\ 2D circuit\end{tabular} & - & $\text{poly}(n)$ & $2^{O(d)} \cdot n^{O(1)}$ \\
%\hline
\cite{vasconcelos2024learning} & \begin{tabular}{l} \QACz\ Unitary,\\$a \leq \mathcal{O} (\log n)$, Frobenius norm\end{tabular} & Proper  & $\mathcal{O}  \br{n^{\log^d n}}$ & $\mathcal{O}  \br{n^{\log^d n}}$  \\
\hline
\cite{vasconcelos2024learning} & \begin{tabular}{l} \QACz\ Unitary, diamond norm\end{tabular} & Improper  & -- & $\Omega(2^n)$  \\
\hline
\cite{ADOY24} & \begin{tabular}{l} \QLCz\ Boolean function \end{tabular}&  Agnostic & $2^{\logO \br{n^{1-3^{-d}}}}$ & $2^{\logO \br{n^{1-3^{-d}}}}$\\
\hline
\textbf{This work} & \begin{tabular}{l} \QLCz\ Boolean function\end{tabular} &  Agnostic & $2^{\logO \br{n^{1-2^{-d}}}}$ & $2^{\logO \br{n^{1-2^{-d}}}}$ \\
\hline
\textbf{This work} & \begin{tabular}{l} \QLCz\ $n \to m$ Channel,\\ Frobenius norm\end{tabular} &  Agnostic & $2^{\logO \br{n^{1-2^{-d}}}}$ & $2^{\logO \br{n^{1-2^{-d}}}}$\\
\hline
\textbf{This work} & \begin{tabular}{l} \QLCz\ $n \to m$ Channel,\\spectral norm of Choi repr.\end{tabular} & Improper  & -- & $\Omega(2^n)$ \\
\hline
\end{tabular}
\caption{Summary of learning results of shallow circuits with many-qubit gates. Here, $a$ means the number of ancilla.  Unless otherwise specified,
the column \textbf{Model} refers to the improper standard PAC model. \QLCz\ means \QACz\ with $a = \mathcal{O}(n)$. And we omit dependencies on $\ve, \delta$ and $m$. }  \label{table: learning-on-shallow-circuits}
\end{table}
\footnotetext[1]{In this table the Frobenius norm refers to the normalized Frobenius norm.}

\paragraph{Property Testing and Tolerant Testing} 
Property testing concerns the task of determining whether a given object satisfies a specified property or is far from satisfying it. It is a central area of research in theoretical computer science. Compared with learning, property testing often admits substantially smaller query complexity. In quantum computing, designing efficient testers for various properties has been studied extensively,~\cite{PhysRevLett.95.260502,harrow2013testing}, including for stabilizer states~\cite{10.1145/3717823.3718277}, junta states~\cite{bao2024learning}, quantum junta channels~\cite{pmlr-v195-bao23b,chen2024tolerant,bao2025efficient} etc. For an overview, see the survey of Montanaro and de Wolf~\cite{montanaro2013survey}.

The standard notion of property testing requires an algorithm to distinguish between objects that exactly satisfy the property and those that are far from it. However, this notion lacks robustness to noise. To address this, Parnas, Ron, and Rubinfeld~\cite{PARNAS20061012} introduced {\em tolerant property testing}, where the goal is to distinguish objects that approximately satisfy the property from those that are far from it. Tolerant testing is particularly significant in the quantum setting, where noise is inherent and unavoidable. Nonetheless, designing efficient tolerant testers is considerably more challenging and, in some cases, provably impossible~\cite{10.1145/3618260.3649687}.

Tolerant testing is also closely related to agnostic learning. In this work, we propose a quantum tolerant tester for determining whether a quantum channel can be implemented by a \QLCz\ circuit building on techniques from agnostic learning.

\subsection{Our Results}

\subsubsection*{Low-degree Concentration and Hardness}

First, similar to \cite{ADOY24}, we bound the approximate degree of the output operator of \QLCz\ channels under low-degree inputs. Specifically, in \cref{sec:approximation}, we prove the following theorem:  

\begin{theorem} [informal of \cref{thm: QAC0-multi-layer}] \label{theorem:informatl-low-degee-concentration}
    Let $U$ be a depth-$d$ \QLCz\ circuit working on $n$ inputs qubits. 
    Let $A$ be an operator with degree $\ell$.
    There exists an operator $M$ with degree $\logO \br{n^{1-2^d} \ell^{-2^d}}$ such that
    \begin{align*}
        \norm{UAU^{\dagger} - M} \leq \ve.
    \end{align*}
\end{theorem}

Our proof leverages the unitary dilation technique to ensure that the approximation of large quantum gates is unitary.  
By embedding the non-unitary, approximated operator into a larger unitary matrix, the technique ensures that the operator can still be managed and analyzed as part of a valid unitary evolution.
This allows us to simultaneously apply the low-degree approximation and light-cone argument to a large generalized Toffoli gate.
This critical property leads to a tighter bound on the approximate degree than the one established in \cite{ADOY24}.

Compared to the $\logO \br{n^{1-3^d} \ell^{-3^d}}$ degree upper bound given by \cite{ADOY24},
our degree upper bound of $M$ is slightly better.
This result directly implies that the Boolean functions computed by \QLCz\ circuits
also have an approximate degree upper bound of $\logO \br{n^{1-2^{-d}}}$,
which matches exactly with the approximate degree upper bound for \LCz\ circuits,
thereby demonstrating the existence of a subexponential time agnostic learning algorithm for  \QLCz\ Boolean functions~\cite{doi:10.1137/1.9781611975482.42}.

Also, as a direct consequence, we can improve the ancilla lower bound and circuit size lower bound for
\QACz\ circuits computing the parity function.
That is, we have
\begin{corollary}[informal of \cref{cor:parity-hardness}]
  Suppose $U$ is a \QACz\ circuit with $n$ input qubits and $a=\tilde{O}\br{n^{1+2^{-d}}}$ ancilla qubits.
  Then $U$ can not approximate the parity function over uniform inputs,
  with a success probability larger than $1/2 + O(d/n)$.
\end{corollary}

\subsubsection*{Agnostic Learning}

We provide an agnostic learning algorithm for \QLCz\ channels.
The main contribution of our work lies in proposing an algorithmic analysis method based on approximation degree. Thus, in \cref{sec:agnostic}, we present the following theorem:

\begin{theorem} [informal of \cref{thm: QLC0-approximation-2-true}]
    There is a subexponential time agnostic learning algorithm for \QLCz\ channels with $n$-qubit input and $\mathrm{polylog}(n)$-qubit outputs with respect to the normalized Frobenius norm.
\end{theorem}

%Noting the close connection between agnostic learning algorithms and tolerant testing problems, we also provide a tolerant testing algorithm for \QLCz\ channels.

Our learning algorithm builds upon the framework of Pauli analysis, which involves learning the Pauli coefficients of the Choi state of the quantum channel. Pauli analysis was introduced by Montanaro and Osborne \cite{MO10}, and has since become a widely used tool in the design of quantum learning algorithms~\cite{doi:10.1137/1.9781611977554.ch43,pmlr-v195-bao23b,NPVY24,arunachalam_et_al:LIPIcs.ICALP.2024.13,Arunachalam2025}. This approach combines with classical shadows \cite{Huang_2020} and Fourier analysis \cite{ODonnell2014}, making it particularly effective in scenarios where the Pauli expansion is sparse—such as juntas and low-degree functions.

We further establish hardness for agnostic learning for \QLCz\ channels
under the spectral norm or diamond norm. More specifically, in \cref{sec:hardness}, we prove an exponential lower bound on learning \QLCz\ channels under the spectral norm or diamond norm.

\begin{theorem} [Hardness of \QACz\ channel learning, informal of \cref{thm: hardness-on-QAC0-few} and \cref{thm: hardness-QAC0-diamond-norm}]
    %Assuming $\C$ is the set of $n \to m$ quantum channels induced by \QACz\ unitaries. 
    %Then,
    Given an unknown $n$-to-$m$ channel $\Phi \in \C$,
    learning the Choi representation of $\Phi$ in spectral norm distance requires $\exp \br{\Omega(n)}$ queries.
    Moreover, if $m= \Omega(n)$,
    then learning $\Phi$ in diamond norm distance requires $\exp \br{\Omega(n)}$ queries.
     
\end{theorem} 

These hardness results also answer a question raised in \cite{wadhwa2024agnostic}, demonstrating that agnostic learning of \QLCz\ channels is impossible under certain stronger norms.

\subsubsection*{Tolerant Testing}

Based on the agnostic learning algorithm for \QLCz\ channels, we also provide a tolerant testing algorithm for \QLCz\ channels. Note that a standard learning algorithm is not sufficient to imply this. The tolerant testing problem requires us to determine whether a channel $X$ is sufficiently close to some \QLCz\ channel or sufficiently far from any \QLCz\ channel.
%The distance from the hypothesis $\tilde{X}$ to the set of \QLCz\ channels can be approximately viewed as the distance from $X$ to the set of \QLCz\ channels. We demonstrate that the entire procedure is feasible by showing that the distance from $\tilde{X}$ to the set of  \QLCz\ channels is also computable.
\begin{theorem}[informal of \cref{cor:tolerant-testing}]
    There exists an $1/\operatorname{poly}(n)$-gap tolerant testing algorithm
    for \QLCz\ channels with $n$-qubit input and $\mathrm{polylog}(n)$-qubit outputs.
    The algorithm has sub-exponential sample and time complexity. 
\end{theorem}

\subsection{Proof Overview}

\paragraph{Low-degree Concentration}

%This is achieved by employing the argument in~\cite{ADOY24} with refinement.
We combine the technique of \textit{unitary dilation} and \textit{operator dilation} with
the low-degree approximation results from \cite{ADOY24}.
The elementary gates in a \QACz\ circuit are the multi-qubit \CZGate s, along with all single-qubit unitaries. The single-qubit unitaries are already of degree $1$, so we do not need further action on them. For large \CZGate\ acting on $n$ qubits, it can be approximated by an operator with Pauli degree at most $\sqrt{n}$, up to logarithmic factors. 

Here comes the dilation technique:
The unitary dilation technique allows us to embed an arbitrary operator with bounded spectral norm into a unitary matrix.
This is especially useful when handling low-degree approximations of large quantum gates,
which lose the property of being a unitary matrix because of the low-degree approximation.
We embed such low-degree approximations into larger unitary operators, and use a layer-by-layer argument to prove an approximate-degree upper bound for \QLCz\ circuits, using similar ideas as in \cite{ADOY24}.

The unitary property allows us to have more refined operations when handling a layer of a \QLCz\ circuit:
Specifically, this allows us to combine a light-cone argument with these low-degree operators:
For a unitary operator $U_i$, we have the identity that $U_i^\dagger U_i = \id$.
That is, for a layer $U = U_1\otimes\cdots\otimes U_s$,
when we consider the Pauli degree of the operator $U^\dagger A U$,
with $A$ being a low-degree operator,
most of the unitary operators $U_i$ will cancel out with themselves.
The result is that the Pauli degree will get multiplied by at most a factor of $\max_i\deg(U_i)$.
With unitary dilation, we can combine this idea with the low-degree approximation of large \CZGate s,
and save a square root factor in this case.

\paragraph{Agnostic Learning for $\QLCz$ circuits in Frobenius norm}
Let $\C$ be the set of Choi states of quantum channels implemented by $\QLCz$ circuits.
We prove that Algorithm \ref{alg: Channel-Tomography} is an agnostic learning algorithm by finding an intermediate class $\mathcal{M}$ such that:
\begin{itemize}
    \item Algorithm \ref{alg: Channel-Tomography} is an agnostic learning algorithm for the concept class $\mathcal{M}$, and,
    \item $\C$ is close to $\mathcal{M}$ in the sense that for any element $C\in \C$,
    there exists an element $M\in\mathcal{M}$ that is close to $C$.
\end{itemize}
We employ Pauli analysis (see \cref{subsec:pauli-analysis}),
and let this intermediate class $\mathcal{M}$ be exactly the operators with a bounded Pauli degree,
which we denote as $\Matrix^{\le d}$ for some $d\in\mathbb{Z}_{\ge 0}$.
The advantage of using operators with a low Pauli degree is that
it is relatively easy to design agnostic learning algorithms with respect to $M^{\le d}$:
For an arbitrary operator $P$ with Pauli decomposition
\begin{equation*}
    P = \sum_{\sigma\in\set{0,1,2,3}^n}\widehat{P}(\sigma)\Base{\sigma},
\end{equation*}
the closest element in $\Matrix^{\le d}$ in the Frobenius norm is exactly the low-degree part of $P$'s Pauli decomposition
\begin{equation*}
    P^{\le d} = \sum_{\substack{\sigma\in\set{0,1,2,3}^n\\\abs{\sigma}\le d}}\widehat{P}(\sigma)\Base{\sigma}.
\end{equation*}
So to agnostically learn an arbitrary operator $P$,
we only need to learn the low-degree part $P^{\le d}$,
which consists of at most $(3n)^d$ Pauli coefficients.
This can be achieved efficiently using classical shadow tomography from Huang, Kueng, and Preskill~{\cite[Lemma 17]{Huang_2020}}.

The remaining task is to prove that any element $C\in\C$ is close to some low-degree operator in $\mathcal{M} = \Matrix^{\le d}$.
This can be achieved by the low-degree concentration results in \cref{theorem:informatl-low-degee-concentration}.

%The final result is that we improve the result \cite{ADOY24} with a constant exponent.

\paragraph{Hardness on Learning $\QLCz$ Channels}

For $n, a \geq 1$ and a \QACz\ circuit $U$,
we say $U$ has a clean computation with $a$ ancilla qubits, if there exists a unitary $V$ such that
\begin{align*}
    U(\ket{\varphi} \otimes \ket{0_a}) = (V\ket{\varphi}) \otimes \ket{0_a}.
\end{align*}
In this case we also call $V$ as a \QACz\ unitary with $a$ ancilla.

Let $\C$ be the set of Choi representations of  $n \to m$ quantum channels implemented by a \QACz\ unitary $V$ with $a$ ancilla. We prove that learning the Choi representation $\Choi{\Phi} \in \C$ under the spectral norm requires an exponential number of queries. 

The proof has two steps. In the first step, we prove that learning $n \to 1$ channels requires an exponential number of queries. In the second step, we generalize the result to $n \to m$ channels.

In the first step, we use a reduction from channel hardness to unitary hardness. 
The main ingredient of the reduction is similar to that in the algorithm \cite[Algorithm 1]{vasconcelos2024learning} provided by Vasconcelos and Huang. 
For a \QACz\ unitary $V$ with $a$ ancilla, suppose $V_i$ is the circuit related to the $i$-th qubit in the sense that $\partrace{-i}{V\rho V^{\dagger}} = \partrace{-i}{V_i\rho V_i^{\dagger}}$. 
Broadly speaking, the beauty of this algorithm lies in the fact that, using the local information of $V_i$, we can sew together the global $V \otimes V^{\dagger}$.
Conversely, as a reduction, if learning the global unitary $V \otimes V^{\dagger}$ is difficult, then learning the local $V_i$ should also be difficult.

To prove the hardness of learning \QACz\ $n \to 1$ channels, we now only need the hardness of learning $V \otimes V^{\dagger}$. This is done with a slightly tailored result from \cite[Proposition 7]{vasconcelos2024learning}.

In the second step, we embed the $n \to 1$ channels into the $n \to m$ channels. We cannot directly use the partial trace because the spectral norm may increase exponentially under this operation. As an alternative, by imposing constraints on the ancilla size, we accomplish this operation by padding $m-1$ irrelevant qubits.

\begin{Anonymous}
\subsection*{Acknowledgement} 
This work was supported by National Natural Science Foundation of China (Grant No. 62332009, and 12347104), Quantum Science and Technology-National Science and Technology Major Project (Grant No. 2021ZD0302901), NSFC/RGC Joint Research Scheme (Grant No. 12461160276), Natural Science Foundation of Jiangsu Province (Grant No. BK20243060) and the New Cornerstone Science Foundation.
\end{Anonymous}

\subsection{Summary and Future Work}\label{sec:future-work}
This paper studies agnostic learning for \QACz\ circuits and their induced channels. We present a subexponential-time agnostic learning algorithm for linear-size \QACz\ $n$-to-$\mathrm{polylog}\br{n}$ channels under the normalized Frobenius norm. Furthermore, we establish that learning such channels under the spectral norm or the diamond norm requires exponential queries.

The field of agnostic learning for quantum states and quantum channels is still in its early stage. Here we list several open problems for future research.

\begin{enumerate}
    \item Is it possible to remove the restriction on the size of the output qubits? Specifically, can we design a subexponential-time agnostic learning algorithm for linear-size \QACz\ unitary operators? Although a PAC learning algorithm exists for this setting -- with subexponential queries and exponential running time -- it remains unclear whether it can be extended to the agnostic case or improved to achieve subexponential runtime.
    % \pnote{move to sec 4}  In fact, our framework guarantees that in algorithm \ref{alg: Unitary-Tomography}, if we have an approximation $V_i S_i V_i^{\dagger}$ under the normalized Frobenius norm, while ensuring that $\norm{Q_i} \leq 1 + \ve_V$, we can derive an algorithm with sub-exponential sample complexity (but exponential time complexity) based on the normalized Frobenius norm. We leave it in the \cref{sec:alg-appendix}. However, how to give an algorithm with sub-exponential time complexity is unknown.

    \item Even subexponential-time algorithms may be too costly for near-term quantum devices. Can we design more efficient agnostic learning algorithms for shallow quantum circuits with geometric structures, such as brickwork or nearest-neighbor architectures?

    \item While learning quantum states prepared by shallow circuits has been studied extensively in the PAC setting, little is known about agnostic learning in this context. Recently, the authors in \cite{10.1145/3717823.3718207}  proposed an agnostic learning algorithm for product states. A natural question is whether similar techniques can be extended to efficiently learn states generated by $\textsf{state-}\QNCz$ or $\textsf{state-}\QLCz$?

    \end{enumerate}

\subsection*{Organization}

In \cref{sec:preliminiries}, we will formally introduce agnostic learning and the \QACz\ model.
In \cref{sec:approximation}, we present improved approximation results for \QACz\ circuits and their applications.
In \cref{sec:agnostic}, we introduce an agnostic learning algorithm for \QACz\ channels with linear ancilla and multiple outputs under the normalized Frobenius norm.
In \cref{sec:tolerant-testing}, we use our agnostic learning algorithm to give a tolerant testing algorithm for \QLCz\ channels.
In \cref{sec:hardness}, we propose a reduction for the hardness of quantum channel learning, thereby establishing corresponding hardness results.
In \cref{sec:future-work}, we discuss some future work.

\section{Preliminaries}\label{sec:preliminiries}

A quantum system $A$ is associated with a finite-dimensional Hilbert space, which we also denote by $A$.
The quantum registers in the quantum system $A$ are represented by {\it density operators},
which are trace-one positive semi-definite operators, in the Hilbert space $A$.
We also use the Dirac notation $\ket{\varphi}$ to represent a pure state.
In this case, we have the convention that $\varphi = \ketbra{\varphi}$,
where here $\varphi$ is a rank-one density operator.
For two separate quantum registers $\varphi$ and $\sigma$ from quantum systems $A$ and $B$,
The compound register is the Kronecker product $\varphi\otimes\sigma$.
A {\it positive operator-valued measure} (POVM) is a quantum measurement described by a set of positive semi-definite operators that sum up to the identity.
Let $\set{P_a}_{a}$ be a POVM applied on a quantum register $\varphi$,
then the probability that the measurement outcome is $a$ is $\Tr\Br{P_a\varphi}$. A quantum channel is a completely positive trace-preserving map. We say  quantum channel is an $n \to m$ channel if it takes $n$-qubit as input and outputs $m$ qubits, i.e., $\Phi:\Matrix_{2^n} \to \Matrix_{2^m}$.

We only consider square matrices in this work.
For any integer $n\geq 2$, let $\mathcal{M}_n$ be the set of $n\times n$ matrices.
The trace of an operator $M\in\Matrix_n$ is $\Tr M = \sum_{i=1}^nM_{i,i}$
and the normalized trace is $\tau M = \frac{1}{n}\sum_{i=1}^nM_{i, i}$.
For any matrix $M\in\mathcal{M}_n$, we let $\abs{M} = \sqrt{M^\dagger M}$. For any $M,N\in\mathcal{M}_n$, the inner product of $M, N$ is $\langle M,N\rangle=\Tr\Br{M^{\dagger}N}/n$. It is evident that $\br{\mathcal{M}_n,\langle\cdot,\cdot\rangle}$ forms a Hilbert space.

For $p\ge 1$, the {\it normalized} Schatten $p$-norm of $M$ is defined to be
\begin{equation*}
    \normsub{M}{p} = \br{\frac{1}{n}\Tr\Br{\abs{M}^p}}^{1/p}.
\end{equation*}
For $p = 2$, it is not hard to see that $\langle M,M\rangle=\normsub{M}{2}^2$. 
Moreover, $\normsub{\cdot}{p}$ is monotone non-decreasing with respect to $p$ and $\normsub{\cdot}{\infty}=\lim_{p\rightarrow\infty}\normsub{\cdot}{p}$ is the spectral norm.  
We often use $\norm{A}$ to refer to $\norm{A}_{\infty}$.

Another norm we need to concern is the Frobenius norm
\begin{align*}
    \norm{M}_F = \sqrt{\sum_{i,j} |M_{ij}|^2} = \br{\Tr\Br{\abs{M}^2}}^{1/2}.
\end{align*}
which is exactly the Schatten $2$-norm.

The following proposition follows from the sub-additivity of spectral norms.

\begin{prop} \label{prop: spectral-norm-for-block-matrix}
    For 
    $A =  \begin{bmatrix}
        A_{11} & A_{12} \\
        A_{21} & A_{22}
    \end{bmatrix}$, 
    $\norm{A} \leq \sum_{ij} \norm{A_{ij}}$.
\end{prop}

\begin{prop} [\Holder\;inequality]
    For operators $A,B$, and $p > 1$,
    \begin{align}
        \norm{AB}_p \leq \norm{A} \cdot \norm{B}_p
    \end{align}
\end{prop}

\begin{prop} [{\cite[Theorem X.1.1]{bhatia2013matrix}}] \label{prop: Lipshitcz-for-sqrt}
     For positive semi-definite operators $A, B$ and $r = 1/2$,
     \begin{align}
         \norm{A^r - B^r} \leq \norm{A-B}^r
     \end{align}
\end{prop}

\begin{definition}[Choi representation and Choi state]\label{def:Choi-representation}
Given a linear map $\Phi : \Matrix_{2^n} \to \Matrix_{2^m}$, its Choi representation is defined as
\begin{align*}
    \Choi{\Phi} = (I \otimes \Phi) ( \ketbra{\EPR{n}} )
\end{align*}
where $\ket{\EPR{n}} =  \sum_{x\in\set{0,1}^n} \ket{x} \otimes \ket{x}$.
%It is easy to see that for a quantum channel $\Phi$, it holds that $\norm{\Choi{\Phi}} \leq 2^m, \Tr \Br{\Choi{\Phi}} = 2^n, \norm{\Choi{\Phi}}_2 \leq 1$
%In this work, we concern ourselves with the approximation of the Choi representation of a quantum channel in terms of  2-norm. 
%When $m$ is small, approximating the Choi representation via the spectral norm is also feasible.
Furthermore, if $\Phi$ is a quantum channel, its Choi state is defined as 
\begin{align}
    \Chois{\Phi} = \frac{1}{2^n}\Choi{\Phi}=\frac{1}{2^n}  (I \otimes \Phi) (\ketbra{\EPR{n}}).
\end{align}
\end{definition}
A unitary $U$ induces a channel $\rho \to U\rho U^{\dagger}$, which we refer to as $\Phi_U$.
We also use $\id$ to refer to the identity channel $\Phi_{\id}$.
For a linear map $\Phi$, its dual map $\Phi^*$ satisfies
\begin{align*}
    \Tr \Br{\Phi(X)Y} = \Tr \Br{X \Phi^*(Y)}.
\end{align*}
Notice that a dual map $\Phi^*$ of a quantum channel $\Phi$ is a quantum channel if and only if $\Phi$ is unital, i.e., $\Phi(\id) = \id$.
The diamond norm $\norm{\Phi}_{\diamond}$ of an $n$-qubit input map is defined as
\begin{equation}\label{eqn:diamondnorm}
\norm{\Phi}_{\diamond} = \max\set{\norm{(\Phi \otimes \id_{2^n})(X)}_1: X\in\Matrix_{2^{2n}}, \norm{X}_1 \leq 1} .
\end{equation}
The following fact holds for unitaries $U$ and $V$, 
\begin{prop} [{\cite[Proposition 1.6]{Haah_2023}}] \label{prop: diamond-is-less-than-spectral}
    Given unitaries $U$ and $V$, we  have
    \begin{align}
    \norm{\Phi_U - \Phi_V}_{\diamond} \leq 2 \norm{U-V}.  
    \end{align}
\end{prop}

For further details regarding distance relations between quantum channels, readers may refer to Yuan and Fung's work \cite{Yuan_2017}.

\subsection{Analysis of Boolean Functions}

In this subsection, we briefly introduce the theory of analysis of Boolean functions. Readers may refer to O'Donnell's excellent book~\cite{ODonnell2014} for a thorough treatment.

Given a Boolean function $f: \set{0,1}^n\to\mathbb{R}$,
its $p$-norm is defined to be $\normsub{f}{p} = \br{\expec{\rx}{\abs{f(\rx)}^p}}^{1/p}$ for $p\geq 1$,
where $\rx$ is a random variable uniformly distributed over $\set{0,1}^n$.
Its infinity norm is defined to be $\normsub{f}{\infty} = \lim_{p\to\infty}\normsub{f}{p} = \max_{x}\abs{f(x)}.$
We will use the notation $\norm{f} = \normsub{f}{\infty}$.
Given Boolean functions $f,g:\set{0,1}^n\to\mathbb{R}$, the inner product of $f$ and $g$ is $\langle f,g\rangle=\expec{\rx}{f(\rx)g(\rx)}$, where $\rx$ is uniformly distributed over $\set{0,1}^n$. 
For any $S\subseteq[n]$, define the Fourier basis $\chi_S$ as
$\chi_S(x) = (-1)^{\sum_{i\in S}x_i}$.
The set $\set{\chi_S}_{S\subseteq[n]}$ is actually an orthonormal basis for Boolean functions.
Any function $f$ admits a Fourier expansion $f = \sum_{S\subseteq[n]}\widehat{f}(S)\chi_S$,
where $\widehat{f}(S)$ are the Fourier coefficients.
Sometimes we consider the case where a Boolean function $f$ has the form $\set{-1,1}^n \to \mathbb{R}$. In this case, $\chi_S(x) = \prod_{i\in S} x_i$.

The following  well-known Parseval's theorem relates the $2$-norm and Fourier coefficients of a Boolean function.
\begin{theorem}[Parseval's theorem]\label{thm:Parseval}
    Let $f: \set{0,1}^n\to\mathbb{R}$ be a Boolean function.
    Then
    \begin{align*}
        \normsub{f}{2}^2 = \sum_{S\subseteq[n]}\widehat{f}(S)^2.
    \end{align*}
\end{theorem}

For Boolean functions taking values in the set $\set{-1,1}$,
one immediately has $\normsub{f}{2} = 1$.

\begin{definition}
    Let $f: \set{0,1}^n\to\mathbb{R}$ be a Boolean function with Fourier expansion
    $$f = \sum_{S\subseteq[n]}\widehat{f}(S)\chi_S.$$
    Then the degree of $f$ is defined as
    \begin{align*}
        \deg\br{f} = \max_{S: \widehat{f}(S) \neq 0} \abs{S}.
    \end{align*}
\end{definition}

% We will use notation $f^{\leq k}$ to refer $\sum_{S: |S| \leq k} \widehat{f}(S)\chi_S$. The notation $f^{<k}, f^{=k}, f^{>k},f^{\geq k}$ are similarly defined. 

\begin{definition}[Approximate Degree]
    Let $f: \set{0,1}^n\to\mathbb{R}$ be a Boolean function.
    For $\varepsilon\in[0, 1]$, the approximate degree of $f$ is defined as
    \begin{align*}
        \degeps{\varepsilon}{f} = \min_{g: \norm{f - g} \le \varepsilon} \deg\br{g}.
    \end{align*}
    %where the notation $\norm{f - g} \le \varepsilon$ means $\abs{f(x) - g(x)} \le \varepsilon$ for each $x\in\set{-1,1}^n$.
    If $\ve$ is not specified, it is $\ve = 1/3$.
\end{definition}

%Generally, when $\ve$ is not mentioned, $\ve = 1/3$. However, in this paper, the value of $\ve$ will always be explicitly specified. 

It is worth noticing that the approximation is with respect to the infinity norm. 

%The notion of approximate degrees has played a crucial role in quantum query complexity and quantum communication complexity~\cite{10.1145/502090.502097,BurhmanWolf:2001}.

% We are particularly interested in the following two classes of Boolean functions.

% \begin{definition}
%     The $\Parityn$ function $\set{-1,1}^n \to \set{-1,1}$ with $n$ inputs is defined as 
%     \begin{align*}
%         \Parityn(x) = \prod_i x_i.
%     \end{align*}
% \end{definition}

% \begin{definition}
%     The $\Majorityn$ function $\set{-1,1}^n \to \set{-1,1}$ with $n$ inputs is defined as 
%     \begin{align*}
%         \Majorityn(x) =  \text{sgn} \br{\sum_i x_i}.
%     \end{align*}
%     where $\text{sgn}(x)=-1$ if $x\geq 0$, and $\text{sgn}(x)=1$ otherwise.
% \end{definition}

% Both functions have large approximate degrees.

% \begin{fact}[{\cite[Example 2.5]{ADOY24}}]
%     For  $1/3> \ve > 2^{-n}$,
%     \begin{align*}
%         \degeps{\ve}{\Parityn} = \Theta\br{n} \\
%         \degeps{\ve}{\Majorityn} = \Theta\br{n}. 
%     \end{align*}
% \end{fact}

\subsection{Pauli analysis}\label{subsec:pauli-analysis}

Pauli analysis is a generalization of the analysis of Boolean functions to the space of matrices $\mathcal{M}_{2^n}$. The Pauli matrices $\Base{0}, \dots, \Base{3}$ are
$$
\Base{0} = \id = \begin{bmatrix}
    1 & 0 \\
    0 & 1
\end{bmatrix},
\Base{1} = \begin{bmatrix}
    0 & 1 \\
    1 & 0
\end{bmatrix},
\Base{2} = \begin{bmatrix}
    0 & -i \\
    i & 0
\end{bmatrix},
\Base{3} = \begin{bmatrix}
    1 & 0 \\
    0 & -1 
\end{bmatrix},
$$
which form an orthonormal basis in $\mathcal{M}_2$ with respect to the inner product $\langle A,B\rangle=\br{\Tr A^{\dagger}B}/2$. For integer $n\geq 1$ and $\sigma\in\set{0,1,2,3}^n$, we define 
$$\Base{\sigma} = \Base{\sigma_1}\otimes\dots\otimes\Base{\sigma_n}.$$
The set of Pauli matrices $\set{\Base{\sigma}}_{\sigma\in\set{0,1,2,3}^n}$ forms an orthonormal basis in $\mathcal{M}_{2^n}$ with respect to the inner product $\langle A,B\rangle=2^{-n}\Tr A^{\dagger}B$.

For a $2^n\times2^n$ matrix $A$,
the Pauli expansion of $A$ is
$$A = \sum_{\sigma\in\set{0,1,2,3}^n}\widehat{A}(\sigma)\cdot\Base{\sigma},$$
where the $\widehat{A}(\sigma)$'s are the Pauli coefficients of $A$.
We can then define the degree and the approximate degree of a matrix in a similar manner:
\begin{definition}\label{def:quantum-approximate-degree}
    Let $n$ be an integer and $A$ be a $2^n\times2^n$ matrix.
    The degree of $A$ is defined as
    $$\deg(A) = \max_{\sigma: \widehat{A}(\sigma) \neq 0} \abs{\sigma},$$
    where $\abs{\sigma} = \abs{\set{i: \sigma_i\neq 0}}$.
    For $\varepsilon\in[0, 1]$, the approximate degree of $A$ is defined as
    $$\degeps{\varepsilon}{A} = \min_{B: \norm{A - B} \le \varepsilon} \deg\br{B},$$
    where $\norm{\cdot}$ is the spectral norm.
\end{definition}

Let $\Matrix_n^{\leq d}$ be the set of $n \times n$ matrices with degree at most $d$. Similar to the classical Parseval's theorem, one can relate the normalized Schatten $2$-norm with the Pauli coefficients:
\begin{theorem}[Parseval's theorem]\label{thm:Parseval_Quantum}
    Let $A \in \Matrix_n$.
    Then
    \begin{equation*}
        \normsub{A}{2}^2 = \sum_{\sigma \in \set{0,1,2,3}^n}\abs{\widehat{A}(\sigma)}^2.
    \end{equation*}
\end{theorem}

We  use the notation $A^{\leq k}$ to refer to $\sum_{\sigma: |\supp{\sigma}| \leq k} \widehat{A}(\sigma)\Base{\sigma}$. The notations $A^{<k}, A^{=k}, A^{>k},A^{\geq k}$ are similarly defined. From the orthogonality of the Pauli basis and Parseval's theorem, we know
\begin{align*}
    \norm{A}_2^2 =  \norm{A^{>k}}_2^2 + \norm{A^{\leq k}}_2^2 .
\end{align*}

\subsection{Agnostic learning}

Kearns, Schapire and Sellie~\cite{10.1145/130385.130424} proposed the agnostic PAC learning model, which is a more general learning model, capturing the case where the learning object may not be in the hypothesis class.
\begin{definition} [Agnostic learning algorithm]
    Let $\mathcal{D}$ be a distribution on $\set{0,1}^n\times\set{0,1}$.
    For any function $h:\set{0,1}^n\rightarrow\set{0,1}$, the error of $h$ relative to $\mathcal{D}$ is defined to be $\mathrm{err}_{\mathcal{D}}\br{h}=\mathrm{Pr}_{\br{x,y}\sim\mathcal{D}}\Br{h\br{x}\neq y}$.
    Let $\C$ be a concept class, which is a class of functions $c: \set{0,1}^n\to\set{0, 1}$,
    One defines $\mathrm{opt}(\C)=\min_{c\in \C}\mathrm{err}_{\mathcal{D}}\br{c}$. 
    
    We say that $\mathcal{A}$ is an $(\ve, \delta)$-agnostic learning algorithm for $\C$ if $\mathcal{A}$ has access to the oracle $\mathcal{D}$, and outputs a hypothesis $h$ with probability at least $1-\delta$ that satisfies 
    \begin{align*}
        \mathrm{err}_{\mathcal{D}}\br{h}\leq \mathrm{opt}(\C)+\varepsilon.
    \end{align*}
    When the hypothesis $h$ we learned satisfies $h \in \C$, we say that the learning algorithm itself is proper. 
\end{definition}

When the support of the distribution $\mathcal{D}$ satisfies that $\forall (x,y) \in \mathcal{D}$, $y = c(x)$ for some $c \in \C$, it falls back to the standard PAC model.
The word ``agnostic'' comes from the unreliability of the access model.

The agnostic model is believed to be closer to the realistic scenario than the standard PAC model,
especially in quantum computing, where noise is unavoidable.
However, designing efficient agnostic learning algorithms is generally challenging. Even very few concept classes are known to be agnostically learnable in subexponential time.  
Bun, Kothari, and Thaler~\cite{doi:10.1137/1.9781611975482.42} gave a subexponential time agnostic learning algorithm for the class of functions with approximate degree $n^c$ for $c<1$, built on~\cite{doi:10.1137/060649057}.

\begin{definition}[Access models]\label{def:access-models}
  When querying a quantum channel, we can have different query models.
  In this work, we consider the following types:
  
  \begin{enumerate}
      \item {\em Choi State Model}.
        For an $n \to m$ quantum channel $\Phi$,
        we are given multiple independent copies of the Choi state $\Chois{\Phi}$ defined in \cref{def:Choi-representation}, which are quantum states on $n+m$ qubits.
      \item {\em Quantum Process Statistical Query (QPSQ) Model}~\cite{Wadhwa_2025}.
        For an $n \to m$ quantum channel $\Phi$,
        in one query we can only obtain classical information about $\Phi$.
        That is, we can access $\Phi$ by first preparing a quantum state $\varphi$ on $n$ qubits,
applying the quantum channel to get the quantum state $\Phi(\varphi)$,
and immediately applying a quantum measurement.
  \end{enumerate}
\end{definition}

In this work, we consider the agnostic learning of quantum channels, which is also referred to as Agnostic Process Tomography introduced by Wadhwa, Lewis, Kashefi and Doosti \cite{wadhwa2024agnostic}.

\begin{definition}[Agnostic Learning of Quantum Channels] \label{def:APT}
    For $\ve, \delta\in(0, 1)$, an algorithm $\mathcal{A}$ is an $(\ve, \delta)$-agnostic learner with respect to a set of quantum channels $\mathscr{L}$ and error function $\err$,
    if given access to an arbitrary channel $\Phi$, the algorithm $\mathcal{A}$ learns a hypothesis $\Phi'$ w.p. $1-\delta$ such that 
    $$
        \err(\Phi, \Phi') \leq \min_{\Phi_C \in \mathscr{L}} \err(\Phi, \Phi_C) + \ve.
    $$
    When the hypothesis $\Phi'$ we learned satisfies $\Phi' \in \mathscr{L}$, we say that the learning algorithm is proper. Otherwise, we say that the algorithm is improper.
\end{definition}
\begin{remark}
  In the improper learning setting, $\Phi'$ is not guaranteed to be a channel.
\end{remark}

\subsection{Error Functions and Distance}

In this work, we assume that the error function $\err$ satisfies the triangle inequality. We mainly focus on the following error distance functions.
Given quantum channels $\Phi$ and $\Psi$, we define the following distances.

\begin{itemize}
\item The {\em normalized Frobenius (norm) distance (of Choi representation)} is $\norm{\Choi{\Phi} - \Choi{\Psi}}_2$.
This distance can be extended to the improper learning setting. Assuming the learned hypothesis for $\Choi{\Phi}$ is $M$, the corresponding error distance is $\norm{\Choi{\Phi} - M}_2.$

\item The {\em spectral (norm) distance (of Choi representation)} is $\norm{\Choi{\Phi} - \Choi{\Psi}}$.
This distance can be extended to the improper learning setting. Assuming the learned hypothesis for $\Choi{\Phi}$ is $M$, the corresponding error distance is $\norm{\Choi{\Phi} - M}.$

\item The {\em diamond (norm) distance} is $\norm{\Phi - \Psi}_{\diamond}$ where the diamond norm is defined in Eq. \eqref{eqn:diamondnorm}. We do not consider the improper learning setting under this error function. 

\end{itemize}

\subsection{Quantum circuit}

A quantum circuit involves an input quantum register initialized with some input state $\ket{\varphi}$,
an ancillary quantum register initialized with some fixed state $\ket{\ancillas}$,
and a series of quantum gates $U_s, \dots, U_1$,
where each $U_i$ is a unitary operator drawn from a predefined gate set $\mathcal{U}$,
and acts on a subset of working qubits.
After the computation, the working quantum registers contain the state
$$U\br{\ket{\varphi}\otimes\ket{\ancillas}} = U_s\dots U_1\br{\ket{\varphi}\otimes\ket{\ancillas}}.$$
%A quantum circuit can be depicted by \cref{fig:quantum-circuit}.

%We can implement an $n$-qubit to $k$-qubit quantum channel
%by tracing out all but the first $k$ qubits of the circuit output.
%This channel with ancilla in the state $\ket{\ancillas}$ is denoted by $\Phi_{k, U, \ket{\ancillas}}$.
%\begin{equation*}
%    \Phi_{k, U, \ket{\ancillas}}(\rho) = ....\dnote{TODO}
%\end{equation*}
%The subscript $k$ is omitted whenever it is clear from the context.
\begin{definition}
    Let $U$ be a unitary implemented by a quantum circuit with $n$ input qubits and $a$ ancilla.
    Let $\ket{\ancillas}$ be a fixed ancilla state on $a$ qubits.
    We use $\Phi_{k, U, \ket{\ancillas}}$ to denote the quantum channel from $n$ qubits to $k$ qubits,
    implemented by $U$ with ancilla $\ket{\ancillas}$, and taking only the first $k$ qubits as output.
    Formally, for any input state $\varphi$ on $n$ qubits, we have
    \begin{equation*}
        \Phi_{k, U, \ket{\ancillas}}(\varphi) = \partrace{\set{k+1, \dots, n+a}}{U\br{\varphi\otimes\ketbra{\ancillas}}U^\dagger}.
    \end{equation*}
    The subscript $k$ is omitted whenever it is clear from the context.
\end{definition}

%\footnote{TODO: ancilla form}

We may get a classical output by applying a computational basis measurement on the first qubit
of the output of a quantum circuit.
That is, we apply the measurement $\set{M_0 = \ketbra{0}\otimes\id, M_1 = \ketbra{1}\otimes\id}$.
With the input state being $\ket{\varphi}$ and the ancillae being $\ket{\ancillas}$,
the probability that we get output $1$ is 
\begin{equation*}
    \Tr\Br{\br{\ketbra{1}\otimes\id}U\br{\ketbra{\varphi}\otimes\ketbra{\ancillas}}U^\dagger}.
\end{equation*}
We use $C_{U, \ket{\ancillas}}$ to denote the above classical output of a quantum channel.
When $\ket{\ancillas} = \ket{0}^a$ or there is no ancilla, we may simply write $C_U$.

% A family of quantum circuits $\set{C_n}_{n\in\mathbb{N}}$ computes a Boolean function $f$ with the worst-case probability $1-\varepsilon$ (with the worst-case error $\varepsilon$) if  for any $x\in\set{0,1}^n$,
%   \begin{equation*}
%     \prob{C_n(x) \neq f(x)} \le \varepsilon,
%   \end{equation*}
%   where $C_n(x)$ is the output of the circuit on input $x$.  
%   Similarly, a quantum circuit computes $f$ with the average-case probability $1-\varepsilon$ (with the average-case error $\varepsilon$) if
%   \begin{equation*}
%     \expec{x\in\set{0,1}^n}{\prob{C_n(x) \neq f(x)}} \le \varepsilon.
%   \end{equation*} 

In this work, we are concerned with \QACz\ circuits, which are polynomial-size constant-depth quantum circuits consisting of single-qubit unitaries and multi-qubit \CZGate s\footnote{Some definitions use generalized Toffoli gates. They are equivalent (by the reduction in \cite{10.5555/2011679.2011682}) in our case.}.
An $n$-qubit \CZGate\ is defined as
\begin{equation}
    \CZG = \id - 2\ketbra{1}^{\otimes n}.
\end{equation}

Thus a \QACz\ circuit implements a unitary $U = L_{d}M_{d}\dots M_1L_0$,
where $d$ is a constant and each $L_i$ is a tensor product of single-qubit unitaries,
and each $M_i$ is a tensor product of \CZGate s.
The depth of this circuit is $d$.

%\QACz\ circuits differ from \ACz\ circuits. In \QACz\ circuits, either single-qubit gates introduce magic \cite{zhang2024unconditional, zhang2025classical} or multi-qubit Toffoli gates introduce long-range correction \cite{bravyi2020quantum}, granting \QACz\ circuits certain counterintuitive capabilities. The other sight arises when \QACz\ circuits are augmented with measurement operations, which similarly enable them to simulate \Fanoutn\ gates \cite{cao2025measurement}.

With a slight abuse of notation, we also use \QACz\ to represent the family of languages that can be decided by  \QACz\ quantum circuits. 
Formally, a language $L$ is in $\QACz$ if there exists a family of constant-depth and polynomial-size quantum circuits $\set{C_n}_{n\in\mathbb{N}}$ consisting of single-qubit gates and polynomial-size CZ-gates, such that for any $n\in\mathbb{N}$ and $x\in\set{0,1}^n$, if $x\in L$ then $\Pr[C_n\br{x}=1]\geq 2/3$, and if $x\notin L$, then $\Pr[C_n\br{x}=0]\geq 2/3$ where $C_n(x)$ is the measurement outcome on the output qubits of the circuit $C_n$ on input $x$. We say a channel is induced by a \QACz\ circuit if the channel can be obtained by implementing a \QACz\ circuit and then tracing out part of the qubits.

We also introduce the class of \QLCz\ circuits, which consists of \QACz\ circuits with linear-size ancilla. \QLCz\ is a quantum counterpart of the classical circuit family \LCz, introduced by Chaudhuri and Radhakrishnan ~\cite{10.1145/237814.237824}, which is one of the most interesting subclasses of \AC[0] and has received significant attention from various perspectives~\cite{1663737,10.1007/978-3-642-11269-0_6,10.1007/978-3-642-31594-7_65}. A quantum channel induced by a \QLCz\ circuit is defined analogously.

\section{Approximation of \texorpdfstring{\QACz}{QAC0} circuits}\label{sec:approximation}

In this section, we use \textit{unitary dilation} combined with the low-degree approximation for \CZGate s,
to prove a new low-degree approximation for \QACz\ circuits, improving the results of Anshu, Dong, Ou, and Yao \cite{ADOY24}.
%Thus, we also obtain a low-degree approximation for the Boolean functions that can be computed by \QACz\ circuits.
%\subsection{Approximation Results}

\begin{restatable}[Low-degree approximation for \QACz\ circuit]{theorem}{ChapterIIImain} 
\label{thm: QAC0-multi-layer}
    Let $n\geq 1$ be an integer.
    Let $U$ be an $n$-qubit unitary implemented by a depth-$d$ \QACz\ circuit.
    Then there exist constants $C_d$ and $C$ such that the following holds.
    
    For any $2^n \times 2^n$ operator $A$ with degree at most $\ell$ and $\norm{A} \leq 1$,
    and any
    \begin{equation}\label{eq:resitriction-on-r}
    r \in \br{2^9 \log n + C_d, \br{n/\ell}^{3^{-1} \cdot 2^{1-d}}} \cup \br{n/\ell, n},
    \end{equation}
    there exists an operator $M$ such that
    \begin{align}
        \norm{UAU^{\dagger} - M} \leq dC n \cdot 2^{-2^{-9}r},
    \end{align}
    and
    \begin{align*}
        \deg\br{M}\leq \mathcal{O}\br{n^{1-2^{-d}}\cdot\ell^{2^{-d}}\cdot r}.
    \end{align*}
\end{restatable}

\begin{remark}
    This result is incomparable to \cite{ADOY24}.
    Although we achieve a better approximation degree upper bound,
    we have the restriction in  \cref{eq:resitriction-on-r} for the error parameter $r$ in this work.
\end{remark}

Note that when the degree of $A$ is upper bounded by $\ell = n^{o(1)}$,
the above theorem holds for any $r=\polylog(n)$.
So we can recover all the results in \cite{ADOY24} with better approximate degree upper bound parameters.
For example, \Cref{thm: QAC0-multi-layer} implies that any \QACz\ circuit that computes \Parityn\ requires at least $a = \tilde{\Omega}\br{n^{1+2^{-d}}}$ ancilla,
slightly improving the $\tilde{\Omega}\br{n^{1+3^{-d}}}$ lower bound in \cite{ADOY24}.
%    The other example is that the approximation degree for Boolean functions computed by 
%    \QLCz\ circuits is $\tilde{O}\br{n^{1-2^{-d}}}$.  This leads to an agnostic learning algorithm for \QACz\ functions with time and sample complexity $2^{\tilde{O}\br{n^{1-2^{-d}}}}$~\cite{10.1145/3444815.3444825}.
\begin{corollary}\label{cor:parity-hardness}
    Suppose $U$ is a \QACz\ circuit with depth $d$,
    and has $n$ input qubits and $a$ ancilla qubits initialized to an arbitrary state.
    Let $C_U: \set{0,1}^n\to\mathbb{R}$ be the function such that $C_U(x)$ is the probability that the circuit $U$
    output $1$ on input $x$.
    Then for the parity function defined as
    $$\Parityn(x) = \bigoplus_ix_i,$$
    for $a = \tilde{O}\br{n^{1+2^{-d}}}$, we have $\prob{C_U(x) = \Parityn(x)} \le \frac{1}{2} + O(d/n)$,
    where $x$ is drawn uniformly at random from the set $\set{0, 1}^n$.
\end{corollary}

\subsection{Unitary Dilation}

We start with the \textit{unitary dilation} technique used in this work,
which is inspired by Montanaro, Shao, and Verdon \cite{montanaro2024lowdegreeapproximationqac0circuits}\footnote{This work is withdrawn by the authors due to some flaws in their proof.}.
%This technique allows us to treat bounded operators as unitary matrices.
The unitary dilation technique allows us to treat non-unitary operators as unitaries
by embedding them into larger unitary operators.
This is especially useful when applying the polynomial methods~\cite{10.1145/28395.28404,tarui1992low} to quantum circuits.
In particular, the low-degree approximations of large quantum gates are not necessarily unitary operators.
By embedding them into larger unitary operators, 
we obtain a unitary approximation with a low-degree submatrix.

\begin{definition} [Unitary Dilation]
Given an operator $A$ with $\norm{A} \leq 1$, its unitary dilation $A^{\uparrow}$ is 
\begin{align}
    A^{\uparrow} = 
    \begin{bmatrix}
        A & (\id-AA^{\dagger})^{1/2} \\
        -(\id-A^{\dagger}A)^{1/2} & -A^{\dagger}
    \end{bmatrix}.
\end{align}
Specifically, for a unitary operator $V$, 
we have $V^{\uparrow} =
    \begin{bmatrix}
        V & 0 \\
        0 & V^{\dagger}
    \end{bmatrix}$.
\end{definition}

The proof of the following fact is deferred to \cref{sec:proofs-appendix}.
\begin{fact}\label{fact:dilation-is-unitary}
    For any operator $A$ with $\norm{A} \le 1$, the operator $A^{\uparrow}$ is a unitary.
\end{fact}

\begin{fact} \label{lemma: spectral-norm-for-dilation}
    For operators $A, B$ with $\norm{A}, \norm{B} \leq 1$ and $\norm{A-B} \leq \ve$, 
    we have 
    \begin{align}
        \norm{A^{\uparrow} - B^{\uparrow}} \leq 5 \sqrt{\ve}.
    \end{align}
\end{fact}

\begin{proof}
    It suffices for us to bound the spectral norm for each submatrix of 
    $A^{\uparrow} - B^{\uparrow}$ with \cref{prop: spectral-norm-for-block-matrix}. 
    
    The top-left and bottom-right parts are simply $A-B$ and $(A-B)^{\dagger}$.
    So they are directly bounded by $\ve$ of the condition.

    Since $\norm{A} \leq 1$, we have $(\id - AA^{\dagger})^{1/2} \geq 0$. 
    Therefore, we can use \cref{prop: Lipshitcz-for-sqrt} to deal with the top-right and bottom-left parts:
    \begin{align*}
        \norm{(\id-AA^{\dagger})^{1/2} - (\id-BB^{\dagger})^{1/2}} 
        \leq \norm{AA^{\dagger} - BB^{\dagger}}^{1/2}.
    \end{align*}
    
    Now, by the triangle inequality,
    \begin{align*}
        \norm{AA^{\dagger} - BB^{\dagger}} \leq 
        \norm{A} \cdot \norm{A^{\dagger} - B^{\dagger}} + \norm{B^{\dagger}} \cdot \norm{A - B} \leq 2 \ve.
    \end{align*}

    Finally, with \cref{prop: spectral-norm-for-block-matrix},
    \begin{align*}
        \norm{A^{\uparrow} - B^{\uparrow}} \leq 2\sqrt{2 \ve} + 2\ve \leq 5\sqrt{\ve}.
    \end{align*}
\end{proof}

We also need the notion of \textit{operator dilation} to incorporate unitary operators obtained from unitary dilation.

\begin{definition} [Operator Dilation]\label{def:operator-dilation}
    Let $n\ge 1$
    and $\mathcal{S} = \set{S_1, \dots, S_m}$ be an ensemble of disjoint subsets of $[n]$.
    For $\sigma\in\set{0,1,2,3}^n$,
    the operator dilation for $\Base{\sigma}$ with respect to the ensemble $\mathcal{S}$ is defined as
    \begin{align}
        \Base{\sigma}^{\Uparrow_\mathcal{S}} := \Base{\sigma} \otimes L(\sigma_{S_1}) \otimes\cdots\otimes L(\sigma_{S_m}),
    \end{align}
    where for any $\tau\in\set{0,1,2,3}^*$, 
    \begin{align}
        L(\tau) = 
        \begin{cases}
            \ketbra{0} & \abs{\tau} \neq 0, \\
            \id & \abs{\tau} = 0.
        \end{cases}
    \end{align}
    For a general operator $A$ with a Pauli expansion $A = \sum_{\sigma} \widehat{A}(\sigma) \Base{\sigma}$, we define
    \begin{align}
        A^{\Uparrow_{\mathcal{S}}} = \sum_{\sigma} \widehat{A}(\sigma) \Base{\sigma}^{\Uparrow_{\mathcal{S}}}.
    \end{align}
    For simplicity, whenever the ensemble $\mathcal{S}$ is irrelevant or clear from context, $\mathcal{S}$ is omitted, and we simply write $A^{\Uparrow}$.
\end{definition}

%In general, we will perform operator dilation simultaneously on $m$ disjoint sets.
%which introduces $m$ new qubits.  To keep the simplicity of the notations, we omit the explicit specification of these sets.

%We can now use \cref{lemma: spectral-norm-for-dilation} to obtain a unitary approximation for \CZGate.
%Not only do we need unitary dilation,  for analytical purposes, but we also need to define a dilation of operators.

An interesting observation is that the operator dilation preserves the spectral norm.

\begin{prop} \label{lemma: spectral-norm-for-operator-dilation}
    Given a matrix $A \in \Matrix_{2^n}$,
    let $A^{\Uparrow}$ be an operator dilation of $A$. Then,
    \begin{align}
        \norm{A^{\Uparrow}} = \norm{A}.
    \end{align}
\end{prop}

\begin{proof}
    Suppose $\opdilate{A}$ is a dilation with respect to the ensemble $\mathcal{S} = \set{S_1, \cdots, S_m}$.
    Then $A^{\Uparrow}$ has the following decomposition:
    \begin{align}
        A^{\Uparrow} &= \sum_{T \subseteq [m]} \ketbra{1_T0_{T^c}} \otimes \sum_{\supp{\sigma} \subseteq \br{\bigcup_{i \in T} S_i}^c} \widehat{A}(\sigma) B_{\sigma}.
    \end{align}
    %This decomposition can be interpreted as follows: positions with ``1" constrain the corresponding entries to be $\id$, while ``0" indicates no such restriction.
    Thus, 
    \begin{align*}
        \norm{A^{\Uparrow}} = 
        \max_T \norm{\sum_{\supp{\sigma} \subseteq \br{\bigcup_{i \in T} S_i}^c} \widehat{A}(\sigma) B_{\sigma}} 
        = \max_T \norm{\partracen{\bigcup_{i \in T} S_i}{A}}.
    \end{align*}
    Then $\norm{A^{\Uparrow}} \geq \norm{A}$ by setting $T = \emptyset$.
    Also, by $\norm{\partracen{S}{A}} \leq \norm{A}$,
    we conclude that $\norm{A^{\Uparrow}} \leq \norm{A}$.
    This concludes the proof.
\end{proof}

\subsection{Proofs of Low-Degree Approximation}

%By applying unitary dilation, we can find a low-degree unitary approximating \CZGate.
In this subsection we prove \cref{thm: QAC0-multi-layer}.
The main idea we use is that a \CZGate\ has a low approximation degree.
\begin{lemma} [Approximation for \CZGate, {\cite[Corollary 3.3]{ADOY24}}] \label{lemma: CZ-approximate-lemma}  
    For integer $n\geq 2$ and real number $1 < r < n$,
    there exists an operator $\widetilde{\CZGr}_n$ such that
    \begin{equation*}
        \norm{\CZG - \widetilde{\CZGr}_n} \le 2^{1-2^{-8}r} \log e,\quad
        \deg\br{\widetilde{\CZGr}_n} \le \sqrt{nr}, \quad\text{and}\quad \norm{\widetilde{\CZGr}_n} \leq 1.
    \end{equation*}
\end{lemma}

With \cref{lemma: spectral-norm-for-dilation} and \cref{lemma: CZ-approximate-lemma}, one has the following corollary.
\begin{corollary} [Dilation of approximation for \CZGate] \label{corollary: CZ-unitary-approximate-lemma}
    For any $n$-qubit \CZGate\ $\CZG$ and real number $1 < r < n$,
    there exists an operator $\widetilde{\CZGr}_n$ such that
    \begin{align}
        \norm{\CZG^{\uparrow} - \widetilde{\CZGr}_n^{\uparrow}} \le \sqrt{10} \cdot 2^{-2^{-9}r} \log e
    \end{align}
    and
    \begin{align*}
        \deg\br{\widetilde{\CZGr}_n} \le \sqrt{nr}.%, \norm{\widetilde{\CZG}} \leq 1.
    \end{align*}
\end{corollary}
    
%We begin with the \QACz\ circuits at the first layer.
We begin by proving low-degree approximations for a single layer of a $\QACz$ circuit.

\begin{lemma} \label{lemma: Qac0-one-layer}
    There exists an absolute constant $C>0$ such that the following holds:
    
    Let $n\ge 1$ be an integer
    and $U = \bigotimes_i\CZGr_{S_i}$ be a layer of \CZGate s,
    where $S_1, \dots, S_m$ are disjoint subsets of $[n]$.
    For each $i$, $\CZGr_{S_i}$ is a $\CZGr_{\abs{S_i}}$ gate acting on qubits in $S_i$.
    For any integer $\ell$ and $r\in (1, \sqrt{n/\ell}) \cup (n/\ell, n)$ 
    and $2^n\times2^n$ operator $A$ with degree at most $\ell$,
    there exists an operator $M$ such that
    \begin{align}\label{eq:one-layer-requirement-1}
        \norm{UAU^{\dagger} - M} \leq Cn \cdot 2^{-2^{-9}r} \cdot \norm{A}
    \end{align}
    and
    \begin{align}\label{eq:one-layer-requirement-2}
        \deg\br{M} \leq 4n^{1/2} \ell^{1/2} r^{1/2}.
    \end{align}
\end{lemma}

The key idea of the proof is as follows.
Let $t$ be a parameter to be specified later.
Divide the \CZGate s into three parts:
\begin{itemize}
    \item For gates of size $\leq t$, we use a lightcone argument.
    \item For gates of size between $t$ and $t^2$, we perform a low-degree approximation and use a lightcone argument.
          To incorporate the lightcone argument with low-degree approximations, we need to use the unitary dilation and operator dilation techniques explained earlier.
    \item For gates of size $> t^2$, we perform a low-degree approximation and use a direct degree argument.
\end{itemize}

The complete proof is deferred to \cref{sec:proofs-appendix}.

We note that, in contrast to \cite{ADOY24}, our approximation requires a range constraint on the parameter $r$. Fortunately, if we require an inverse-polynomial error,
we only need $r = O(\log n)$ and this remains within the applicable scope of the lemma. 

Single-qubit unitaries do not change the degree of an operator. 
So, after applying \cref{lemma: Qac0-one-layer} for $d$ times, we obtain \cref{thm: QAC0-multi-layer},
which we rephrase below for the reader's convenience.

\ChapterIIImain*
% \begin{theorem} \label{thm: QAC0-multi-layer}
%     Let $n\geq 1$ be an integer and $U$ be a depth $d$ \QACz\ circuit. There exists constants $C_d$ and $C$ such that the following holds.
    
%     For any $r \in (C_d \log n, \br{n/\ell}^{3^{-1} \cdot 2^{1-d}}) \cup (n/\ell, n)$ and any $2^n \times 2^n$ operator $A$ with degree at most $\ell$ and $\norm{A} \leq 1$,
%     there exists an operator $M$ such that
%     \begin{align}
%         \norm{UAU^{\dagger} - M} \leq dC n \cdot 2^{-2^{-9}r}.
%     \end{align}
%     and
%     \begin{align}
%         \deg\br{M}\leq \mathcal{O}\br{n^{1-2^{-d}} \ell^{2^{-d}}r}.
%     \end{align}
% \end{theorem}

\begin{proof}
    Let $p = \tilde{C}n \cdot 2^{-2^{-9}r}$ where $\tilde{C}$ is the constant in \cref{lemma: Qac0-one-layer}. 
    Choose $C = 2\tilde{C}$ and $C_d$ large enough such that $\tilde{C}2^{-2^{-9}C_d} \leq 2^{1/d} - 1$. 
    The constant selection ensures that $(1+p)^d \leq 2$.
    
    Let $U$ be $U = V_d L_d V_{d-1} L_{d-1} \cdots V_1L_1$ where $V$ stands for multi-qubit unitary layer and $L$ stands for single-qubit unitary layer. Write $U_{\leq 0} = U_0 = \id, U_{i} = V_i L_i$ and $ U_{\leq i} = U_{i} U_{\leq i - 1}$.
    
    We define operators $\set{M_i}$ inductively for $i=0, \dots, d$. Set $M_0 = A$.

    We will prove by induction that for any $i=0, \dots, d-1$, we have
    \begin{equation}\label{eq:approximate-main-induction-1}
        \norm{U_{\leq i} A U_{\leq i}^{\dagger} - M_i} \leq iCn \cdot 2^{-2^{-9}r}
    \end{equation}
    and
    \begin{equation}\label{eq:approximate-main-induction-2}
        \norm{M_i} \leq (1+p)^i,\quad \deg \br{M_i} \leq \mathcal{O}\br{n^{1-2^{-i}} \ell^{2^{-i}}r}.
    \end{equation}

    The base case is $i=0$.
    Indeed, for \cref{eq:approximate-main-induction-1} it is easy to check that
    \begin{equation*}
        \norm{U_{\leq 0} A U_{\leq 0}^{\dagger} - M_0} = \norm{A - A} = 0.
    \end{equation*}
    and \cref{eq:approximate-main-induction-2} holds for $i=0$.
    Now suppose \cref{eq:approximate-main-induction-1} and \cref{eq:approximate-main-induction-2} hold for some $i$.
    We carefully choose $r$ such that \cref{lemma: Qac0-one-layer} remains applicable to
    the parameters $A\gets M_i$ and $U\gets U_i$,
    and one can find $M_{i+1}$ such that 
    \begin{align}
        \norm{U_i M_i U_i^{\dagger} - M_{i+1}} \leq \tilde{C}n2^{-2^{-9}r} \cdot \norm{M_i}
    \end{align}
    and
    \begin{align}
        \deg \br{M_{i+1}} 
        \leq \mathcal{O}\br{ \br{n^{1-2^{-{i}}} \ell^{2^{-{i}}}r}^{1/2} r^{1/2}}
        = \mathcal{O}\br{n^{1-2^{-{i+1}}} \ell^{2^{-{i+1}}}r}.
    \end{align}
    Recall that $\norm{M_i} = (1+p)^i \leq (1+p)^d \leq 2$.
    Using the triangle inequality, we now bound the spectral norm:
    \begin{align*}
        \norm{U_{\leq i+1} AU_{\leq i+1}^{\dagger} - M_{i+1}} &=
        \norm{U_{\leq i}AU_{\leq i}^{\dagger} - U_{i+1}^{\dagger}M_{i+1}U_{i+1}} \\
        &\leq \norm{U_{\leq i}AU_{\leq i}^{\dagger} - M_i} + \norm{M_i - U_{i+1}^{\dagger}M_{i+1}U_{i+1}} \\
        &\leq iCn2^{-2^{-9}r} + 2\tilde{C}n2^{-2^{-9}r} = (i+1)Cn2^{-2^{-9}r}.
    \end{align*}
    We finish our induction process by estimating the norm of $M_{i+1}$.
    \begin{align*}
        &\;\; \norm{U_i M_i U_i^{\dagger} - M_{i+1}} \leq \tilde{C}n2^{-2^{-9}r} \cdot \norm{M_i} \\
        &\Rightarrow \norm{M_{i+1}} \leq (1+p)\cdot \norm{M_i} \leq (1+p)^{i+1}.
    \end{align*}

    Taking $M$ as $M_d$  completes the proof.
\end{proof}

\section{Agnostic Learning for \texorpdfstring{\QLCz}{QLC0} channels}\label{sec:agnostic}

%In this section, we present the agnostic learning algorithms for \QLCz\ channels.  
%The correctness proof of this algorithm relies on the low-degree approximation of the \QLCz\ circuit. Since both this paper and \cite{ADOY24} provide low-degree approximations in incomparable senses, we present two learning algorithms in \cref{thm: QLC0-approximation-2-true} and \cref{thm: QAC0-agnostic-algorithm-two} with different parameters. 
%The difference is that: 
%if we require the same error, \cref{thm: QLC0-approximation-2-true} is more efficient (with a sub-exponential speed up). 
%But, \cref{thm: QAC0-agnostic-algorithm-two} is more robust: it works regardless error while \cref{thm: QLC0-approximation-2-true} has a requirement on the error.

%Suppose $\Phi: \Matrix_n\to\Matrix_m$ is a quantum channel implemented by a \QACz\ circuit using $a$ additional ancilla.

%\dnote{First we explain this sentence. TODO}
%Thus, we provide two agnostic learning algorithms .

In this section, we present an agnostic learning algorithm for $\QLCz$ channels with respect to the Frobenius norm as in Algorithm \ref{alg: Channel-Tomography}.
This algorithm employs the classical shadow tomography introduced by Huang, Kueng and Preskill~\cite{Huang_2020}.

\begin{lemma}[Classical Shadows for Low-degree Pauli Operators, {\cite[Lemma 17]{Huang_2020}}] \label{thm:Classical_Shadows}
    For any $\ve, \delta\in(0, 1)$, there exists an algorithm such that:
    
    Given a quantum state $\rho$ and a list of observables $\set{O_1,\cdots,O_M}$, satisfying $\norm{O_i} \leq 1$  for all $i \in [M]$,
    the algorithm outputs estimates $\set{\hat{o}_i}$ satisfying
\begin{align}
    \abs{\hat{o}_i - \Tr \br{O_i \rho}} \leq \ve, \forall i \in [M]
\end{align}
with probability at least $1 - \delta$.

When the $O_i$ are degree-$d$ Pauli operators, one needs
\begin{align*}
    N = \mathcal{O} \br{ \frac{3^d \log(M/\delta)}{\ve^2} }
\end{align*} 
copies of $\rho$, and the (classical) computation time is $\mathcal{O}(dNM).$
%\pnote{what does "classical computation  performed in time" mean?}

\end{lemma}

% This is a standard algorithm
% which has already been employed in \cite{wadhwa2024agnostic, arunachalam_et_al:LIPIcs.ICALP.2024.13, NPVY24}.
%The  algorithm \ref{alg: Channel-Tomography} details on how to learn the channel in the sense of the 2-norm. This is a standard technique that has also been employed in \cite{wadhwa2024agnostic}, \cite{arunachalam_et_al:LIPIcs.ICALP.2024.13} and \cite{NPVY24}.
%We now demonstrate that it is in fact an agnostic algorithm.

\begin{algorithm}[htp]
\SetKwInOut{Input}{Input}
\SetKwInOut{Output}{Output}
\SetKwInOut{Parameters}{Parameters}

\caption{\algname{Channel-Learning}($\Phi$, $d$, $\delta$, $\ve$)}
\label{alg: Channel-Tomography}
\Parameters{$d\in\mathcal{Z}_{\ge 0}$, and $\ve, \delta\in(0, 1)$.}
\Input{Access to the Choi state $\Chois{\Phi}$ of an $n \to m$ quantum channel $\Phi$, where the Choi state of a quantum channel is defined in \cref{def:Choi-representation}}
\Output{An approximation $L$ such that $\norm{L - \br{\Choi{\Phi}}^{\leq d}}_2 \leq \ve$ with probability $1-\delta$
%of the Choi representation $\Choi{\Phi}$
}

%\begin{algorithmic}[1]
%\item For any $\sigma\in\set{0,1,2,3}^{n+m}$ with $|\supp{\sigma}| \leq d$, learn $\hat{\alpha}_{\sigma} \approx \widehat{\Choi{\Phi}}(\sigma)$ with the Classical Shadow Tomography algorithm (\cref{thm:Classical_Shadows}):
%    \begin{itemize}
%        %\item Let $O_i$ be the $i$-th Pauli operator $\Base{\sigma}$ with $|\supp{\sigma}| \leq d$.
%        \item Perform Classical Shadow Tomography on observables $\set{\Base{\sigma}: \abs{\sigma}\le d}$. 
%        \item For the $i$-th Pauli operator $\Base{\sigma}$ with $|\supp{\sigma}| \leq d$, set $\hat{\alpha}_{\sigma} = \hat{o}_i \cdot 2^{-m}$.
%    \end{itemize}
%\item Return $L = \sum_{\sigma: |\supp{\sigma}| \leq d} \hat{\alpha}_{\sigma} \Base{\sigma}$.
%\end{algorithmic}

\begin{algorithmic}[1]
  \item Perform the classical shadow tomography algorithm in \cref{thm:Classical_Shadows}
    with parameters $\ve  \gets \frac{\ve^2}{d(n+m)^{d} 2^m}$ and $\delta \gets \delta$,
    for the quantum state $\Chois{\Phi}$ with Pauli observables $\set{\Base{\sigma}: \abs{\sigma} \le d}$,
    to get a list of estimates $\set{\hat{o}_{\sigma}: \abs{\sigma} \le d}$.
  \item For each $\sigma$ with $\abs{\sigma} \le d$,
    set $\hat{\alpha}_\sigma = \hat{o}_{\sigma} \cdot 2^{-m}$.
  \item Output $L = \sum_{\sigma: \abs{\sigma} \leq d} \hat{\alpha}_{\sigma} \Base{\sigma}$.
\end{algorithmic}

\end{algorithm}

%In subsection 4.1, we introduce the Classical Shadow method and the learning approach based on it. In subsection 4.2, we demonstrate the learning effectiveness of this method. Finally, in subsection 4.3, we discuss some details of the algorithm and its potential applications.

Our main result shows that for any $\ve \ge 1/\poly(n)$ and $\delta \in (0, 1)$,
Algorithm \ref{alg: Channel-Tomography} is an $(\ve, \delta)$-agnostic learner
with respect to the class of quantum channels implemented by $\QLCz$ circuits,
using only sub-exponential queries.
Formally,
\begin{theorem} \label{thm: QLC0-approximation-2-true}
    Given $0 < \ve,\delta < 1$ and $n,m,d,a \geq 1$, and
    assume $\C$ is a set of Choi representations of $n \to m$ quantum channels
    implemented by depth-$d$ \QACz\ circuits with $a = \mathrm{poly}(n)$ ancilla.
    For any $\ve$ satisfying
    \begin{equation*}
        \ve \ge -\exp\br{n^{2^{-d-2}}/m},
    \end{equation*}
    % defined in \cref{thm: QLC0-approximation-2}
    let
    \begin{align*}
        D = \logO \br{ (n+a)^{1-2^{-d}} \cdot m^{2^{-d} + 1}  \cdot \log^{2}(1/\ve) },
    \end{align*}
    where $\logO$ hides log factors.
    Algorithm \ref{alg: Channel-Tomography} with parameters $(d, \ve, \delta) \gets (D, \ve, \delta)$
    is an $(\ve, \delta)$-agnostic learner with respect to $\C$ and error function $\err(X,Y) = \norm{X-Y}_2$.
    %taking
    %\begin{align*}
    %    \mathcal{O} \br{\frac{4^m (n+m)^{2D} \log \left( \frac{(n+m)^D}{\delta}\right)}{\ve^4}}
    %\end{align*}
    The number of queries to $\Phi$ and the running time of the algorithm are both 
    \begin{align*}
         \mathcal{O} \br{\operatorname{poly}(n,m) \cdot 4^m\cdot(n+m)^{3D}\cdot\frac{\log(1/\delta)}{\ve^4}}.
    \end{align*}
    %\begin{align*}
    %     \mathcal{O}  \br{ \frac{4^m D^3 (n+m)^{3D}\log \left( \frac{(n+m)^D}{\delta}\right)}{\ve^4}}.
    %\end{align*}
      In particular, when $m=\mathrm{polylog}(n), \ve = 1/\mathrm{poly}(n)$, and $a=\mathcal{O}(n)$, the number of queries to $\Phi$ and the running time are both $2^{\tilde{O} \br{n^{1-2^{-d}}}}$.
\end{theorem}

%\begin{remark}
%    Note that in algorithm \ref{alg: Channel-Tomography}, we use the Choi-state based access model (that is, given the access to the Choi state of a channel).
%    As discussed in Section 4.3 of \cite{NPVY24}, while the Choi-state based access model is analytically convenient, it is difficult to implement in real physical environments. A more natural model is the Quantum Process Statistical Query (QPSQ) model proposed in \cite{Wadhwa_2025}, where, given a channel $\Phi$, one prepares an input state 
%    $\ket{\varphi}$, applies $\Phi$, and performs measurements. \pnote{maybe include the def in the preliminary, and state it as a corollary}
%    Under QPSQ model, we can still have an algorithm, but at the cost of increasing the number of samples with a factor $(n + m)^D$.
%\end{remark}

We can also get a variant of \cref{thm: QLC0-approximation-2-true} where we don't have the restriction on $\ve \ge 1/\poly(n)$.
For technical reasons, however, we need to slightly increase the number of queries.

\begin{theorem} \label{thm: QAC0-agnostic-algorithm-two}
    Given $0 < \ve,\delta < 1$ and $n,m,d,a \geq 1$, and
    assume $\C$ is a set of Choi representations of $n \to m$ quantum channels
    implemented by depth-$d$ \QACz\ circuits with $a = \mathrm{poly}(n)$ ancilla. 
    % QACz, a = poly(n)
    For
    \begin{align*}
        D = \logO \br{ (n+a)^{1-3^{-d}} \cdot m^{3^{-d} + 1/2}  \cdot \log(1/\ve) },
    \end{align*} 
    Algorithm \ref{alg: Channel-Tomography} with parameters $(d, \ve, \delta)\gets(D, \ve, \delta)$,
    is an $(\ve, \delta)$-agnostic learner with respect to $\C$ and error function $\err(X,Y) = \norm{X-Y}_2$.
    %with
    %\begin{align*}
    %    \mathcal{O} \br{\frac{4^m (n+m)^{2D} \log \left( \frac{(n+m)^D}{\delta}\right)}{\ve^4}}
    %\end{align*}
    The number of queries to $\Phi$ and the running time are both
    \begin{align*}
         \mathcal{O} \br{\operatorname{poly}(n,m) \cdot 4^m\cdot(n+m)^{3D}\cdot\frac{\log(1/\delta)}{\ve^5}}.
    \end{align*}
    %\begin{align*}
    %     \mathcal{O}  \br{ \frac{4^m D^3 (n+m)^{3D}\log \left( \frac{(n+m)^D}{\delta}\right)}{\ve^4}}.
    %\end{align*}
  In particular, when $m=\mathrm{polylog}(n), \ve = 1/\mathrm{poly}(n)$, and $a=\mathcal{O}(n)$, the number of queries to $\Phi$ and the running time are both
  $2^{\tilde{O} \br{n^{1-3^{-d}}}}. $
\end{theorem}

%\begin{remark}
%\dnote{What is the purpose of this remark?}
%    We focus on the case when the algorithm in \cref{thm: QAC0-agnostic-algorithm-two} is sub-exponential:
%    
%    Assuming $\C$ is a set of quantum channels induced by \QACz\ circuits with $n$ inputs, 
%    $\Theta(n)$ ancilla, depth-$d$ and taking the first quasi-polylog qubits as outputs. Then, there is an $(\exp(-o(n)), \exp(-o(n)))$-agnostic learner with respect to $\C$ and error function $\err(X,Y) = \norm{X-Y}_2$ with sub-exponential query complexity and computational complexity.
%\end{remark}

Note that the above theorems assume the Choi state query model.
While the Choi-state access model is analytically convenient, it isn't easy to implement in real physical environments,
where a more natural model is the  Quantum Process Statistical Query (QPSQ) model.\footnote{See \cref{def:access-models} for the definitions of these query models.}
%As discussed in Section 4.3 of \cite{NPVY24}, under the QPSQ model, the condition (1) in \cref{lemma: Learning_Skeleton} still holds at the cost of increasing the number of samples with a factor $(n + m)^D$.
Luckily, in this model we can use \cite[Lemma 40]{NPVY24} to achieve the same effect of the shadow tomography algorithm in \cref{thm:Classical_Shadows}, with a $(n+m)^D$ number of queries overhead.
Thus, in the QPSQ model, we also have an agnostic learning algorithm, but at the cost of increasing the number of samples by a factor of $(n + m)^D$.

\begin{corollary}\label{cor:agnostic-tomo}
    Given $0 < \ve,\delta < 1$, $n,m,d \geq 1$ and $a \geq 0$, and
    assume $\C$ is a set of Choi representations of $n \to m$  quantum channels implemented by depth-$d$ \QACz\ circuits with $a$ ancilla. 
    For
    \begin{align*}
        D = \logO \br{ (n+a)^{1-3^{-d}} \cdot m^{3^{-d} + 1/2}  \cdot \log(1/\ve) },
    \end{align*} 
    Under the QPSQ access model, there is an $(\ve, \delta)$-agnostic learner with respect to $\C$ and error function $\err(X,Y) = \norm{X-Y}_2$, 
    The number of queries to $\Phi$ and the computation time are both
    \begin{align*}
         \mathcal{O} \br{\operatorname{poly}(n,m) \cdot 4^m\cdot(n+m)^{4D}\cdot\frac{\log(1/\delta)}{\ve^5}}.
    \end{align*}
    %\begin{align*}
    %     \mathcal{O}  \br{ \frac{4^m D^3 (n+m)^{4D}\log \left( \frac{(n+m)^D}{\delta}\right)}{\ve^4}}.
    %\end{align*}
\end{corollary}

\subsection{Analysis of Agnostic Learning Algorithm}
We start by analyzing the algorithm in \cref{thm: QLC0-approximation-2-true}.
The proof of \cref{thm: QAC0-agnostic-algorithm-two} is similar,
where we will simply replace \cref{thm: QLC0-approximation-2} with \cref{prop: QLC0_approximate_lemma_1}.

%The following lemma demonstrates that for any family of channels,
%providing an agnostic tomography algorithm can be reduced to verifying two conditions:

We prove that Algorithm \ref{alg: Channel-Tomography} is an agnostic learning algorithm
by finding an intermediate set of objects $\mathcal{M}$ that is close to the concept class $\C$,
and at the same time, Algorithm \ref{alg: Channel-Tomography}
is also an agnostic learning algorithm with respect to $\mathcal{M}$.
In our work, this intermediate set $\mathcal{M}$ is the low-degree operators.
\begin{lemma}[Skeleton of Agnostic Learning] \label{lemma: Learning_Skeleton}
    Given $0 < \ve_1,\ve_2,\delta < 1$ and $n,m \geq 1$, and
    assume $\C$ is a set of Choi representations of $n \to m$ channels,
    and $\mathcal{A}$ is a learning algorithm.
    Recall $\Matrix_{2^{n+m}}^{\leq d}$ is the set of matrices with Pauli degree at most $d$.
    For simplicity, denote $\Matrix^{\leq d} = \Matrix_{2^{n+m}}^{\leq d}$.
    Suppose the following two conditions hold for some error function $\err$ on matrices satisfying the triangle inequality.
    \begin{enumerate}[(1)]
        \item Algorithm $\mathcal{A}$ is an agnostic learning algorithm with respect to the concept class $\Matrix^{\le d}$:
        There exists $\ve_1 > 0$ such that,
        given access to any channel $\Phi$,
        the algorithm $\mathcal{A}$ outputs an $L\in\Matrix_{2^{n+m}}$ in time $T(n,\delta,\ve_1)$
        such that with probability at least $1-\delta$,
        \begin{equation}\label{eq:c95c4528-bf86-439b-823b-194d821b750f}
         \err(\Choi{\Phi}, L) \leq \min_{M \in \Matrix^{\leq d}} \err(\Choi{\Phi}, M) +\ve_1.
        \end{equation}
        \item The class $\C$ is close to $\Matrix^{\le d}$:
        Any Choi representation in $\C$ can be approximated by some matrix in $\Matrix^{\leq d}$.
        That is, there exists $\ve_2 > 0$ such that for any $\Choi{\Phi} \in\C$,
        \begin{equation}\label{eq:0ef69b5d-ff9b-4475-9d0e-1650fc1b8dd5}
            \min_{M \in \Matrix^{\leq d}} \err(\Choi{\Phi} , M) \leq \ve_2.
        \end{equation}
    \end{enumerate}
    Then $\mathcal{A}$ is an $(\ve_1 + \ve_2, \delta)$-agnostic learner for $\C$ in time $T(n,\delta,\ve_1)$ with respect to error function $\err$.
\end{lemma}

\begin{proof}
    Fix any $n \to m$ channel $\Phi$ to be learned.
    Let $\Choi{\Psi^*}  \in \C$ be the minimizer of
    \begin{align*}
        \min_{\Choi{\Psi}  \in \C} \err(\Choi{\Phi}, \Choi{\Psi}).
    \end{align*}
    Let $\text{opt} = \err(\Choi{\Phi}, \Choi{\Psi^*} )$.
    %recall that we only consider the error function satisfying triangle inequality.
    By the triangle inequality and then \cref{eq:0ef69b5d-ff9b-4475-9d0e-1650fc1b8dd5},
\begin{align*}
    \min_{M \in \Matrix^{\leq d}} \err(\Choi{\Phi}, M) 
    &\leq \err(\Choi{\Phi}, \Choi{\Psi^*}) + \min_{M \in \Matrix^{\leq d}} \err(\Choi{\Psi^*}, M) \\
    &\leq \text{opt} + \ve_2.
\end{align*}
%\pnote{Although $\err(\cdot)$ is not specified, does it to satisfy triangle inequality?}
    Suppose algorithm $\mathcal{A}$ outputs $\tilde{M}$ as an approximation  of $\Choi{\Phi}$.
    By \cref{eq:c95c4528-bf86-439b-823b-194d821b750f},
\begin{align*}
    \err(\Choi{\Phi},\tilde{M}) & \leq \min_{M \in \Matrix^{\leq d}} \err(\Choi{\Phi}, M)  + \ve_1 \\
    & \leq \text{opt} + \ve_1 + \ve_2.
\end{align*}
\end{proof}

From now on, choose $\operatorname{err}(X,Y) = \norm{X-Y}_2$. 
For Algorithm \ref{alg: Channel-Tomography}, we verify the two conditions in \cref{lemma: Learning_Skeleton}, separately by \cref{lemma: approx-channel-2-norm} and \cref{thm: QLC0-approximation-2}.

We first check the condition (1) that Algorithm \ref{alg: Channel-Tomography}
is an agnostic learning algorithm with respect to the class $\Matrix^{\le d}$.

\begin{lemma} \label{lemma: approx-channel-2-norm}
    For any parameters $1 \leq d \leq n + m$, and $0 < \ve, \delta < 1$,
    Algorithm \ref{alg: Channel-Tomography} satisfies condition (1) in \cref{lemma: Learning_Skeleton} with $\ve_1\gets\ve$.
    That is, given access to an $n \to m$ channel $\Phi$,
    Algorithm \ref{alg: Channel-Tomography} outputs an $L\in\Matrix_{2^{n+m}}$ such that with probability $1-\delta$,
    \begin{align}
        \norm{L - \Choi{\Phi}}_2 \leq \min_{M \in \Matrix_{2^{n+m}}^{\leq d}} \norm{M - \Choi{\Phi}} +\ve.
    \end{align}
    Moreover, Algorithm \ref{alg: Channel-Tomography} uses
    \begin{align*}
        \mathcal{O} \br{\frac{4^m (n+m)^{2d} \log \br{ \frac{(n+m)^d}{\delta} } }{\ve^4}}
    \end{align*}
    queries to $\Chois{\Phi}$ and the computational time is 
    \begin{align*}
         \mathcal{O} \br{\frac{4^m d^3 (n+m)^{3d}\log \br{ \frac{(n+m)^d}{\delta} } }{\ve^4}}.
    \end{align*}
\end{lemma}

\begin{proof}[Proof of Lemma \ref{lemma: approx-channel-2-norm}]
    Estimating Pauli coefficients of the Choi representation can be finished by applying classical shadow to the Choi state. This fact also indicates that in Algorithm \ref{alg: Channel-Tomography}, classical shadow tomography is essentially estimating the Pauli coefficients.
    \begin{align}
        \Tr \Br{\Chois{\Phi} \Base{\sigma}} =  \frac{1}{2^n} \cdot 2^{n+m} \cdot \widehat{\Choi{\Phi}}(\sigma) = 2^m \widehat{\Choi{\Phi}}(\sigma).
    \end{align}
    
    For simplicity, define $\alpha_{\sigma} = \widehat{\Choi{\Phi}}(\sigma)$. 
    Now, using Theorem \ref{thm:Classical_Shadows} with argument $\tilde{\ve} = \frac{\ve^2}{d(n+m)^{d} 2^m}$, we know that the approximation results $\set{\hat{\alpha}_{\sigma}}$ satisfy
    \begin{align*}
        \abs{\hat{\alpha}_{\sigma} - \alpha_{\sigma}} \leq \tilde{\ve}.
    \end{align*}

    Due to Theorem \ref{thm:Parseval_Quantum},
    \begin{align}
        \norm{L - \Choi{\Phi}^{\leq d}}_2 \leq \sqrt{d(n+m)^{d} 2^m \tilde{\ve}} = \ve.
    \end{align}
        Let $M^* \in \Matrix_{2^{n+m}}^{\leq d}$ be the matrix minimizing $\norm{\Choi{\Phi} - M^*}_2$,
        in fact, we have
    \begin{align*}
        M^* = \sum_{\sigma: |\supp{\sigma}| \leq d} \alpha_{\sigma} \Base{\sigma} = \Choi{\Phi}^{\leq d}.
    \end{align*}
    Thus,
    \begin{align*}
        \norm{\Choi{\Phi} - L}_2
        &\leq \norm{\Choi{\Phi} - M^*}_2 + \norm{L - M^*}_2 \\
        &\leq \norm{\Choi{\Phi} - M^*}_2 + \ve \\
        &= \min_{M \in \Matrix_{2^{n+m}}^{\leq d}} \norm{\Choi{\Phi} - M}_2 + \ve.
    \end{align*}

\end{proof}

% \begin{lemma} \label{lemma: well-approximation-2-norm}
%     Suppose $\C$ is a set of Choi representations of channels with the following structure:
%     Given a depth-$d$ \QACz\ circuit with $n$ input qubits and $a = \mathrm{poly}(n)$ ancilla initialized in the pure state $\psi$.  For $m \leq n$ where $m \cdot \log(1/\ve_1) < n^{C_d} $ for some constant $C_d$ depending only on $d$, defined in \cref{def: C_d in algorithm}, take the first $m(n)$ qubits as outputs and implement the quantum channel
%     \begin{align}
%          \mathcal{E}_{m, U,\psi}(\rho) = \partrace{[m]^c}{U(\rho \otimes \psi) U^{\dagger}}.
%     \end{align}

    % For $D$ with
    % \begin{align*}
    %     D \leq \tilde{ \mathcal{O}} 
    %      \br{ (n+a)^{1-2^{-d}} m^{1+2^{-(d-1)}}  \log^{1+2^{-d}}(1/\ve_1) },
    % \end{align*}
    % $\C$ satisfy condition (2) in \cref{lemma: Learning_Skeleton}. I.e., for any $C \in \C$, we have
    % \begin{align}
    %     \min_{M \in \Matrix^{\leq D}} \norm{M-C}_2 \leq \ve_1.
    % \end{align}
% \end{lemma}

Now, let us turn to the task of verifying condition (2) in \cref{lemma: Learning_Skeleton}.
We use the following proposition, which is based on the low-degree approximation results of \QLCz\ circuits.

%Below we present the lemmas needed to prove \cref{lemma: well-approximation-2-norm}.

\begin{prop} \label{thm: QLC0-approximation-2}
    Given $0 < \ve < 1$, $n,m,d \geq 1$ satisfying $m \cdot \log(1/\ve) < n^{2^{-d-2}}$,
     $a \geq 0$, and
    assume $\C$ is a set of Choi representations of $n \to m$ quantum channels implemented by depth-$d$ \QACz\ circuits with $a = \mathrm{poly}(n)$ ancilla initialized in a pure state $\psi$. 
    % and implement the quantum channel
    % \begin{align*}
    %      \mathcal{E}_{m, U,\psi}(\rho) = \partrace{[m]^c}{U(\rho \otimes \psi) U^{\dagger}}.
    % \end{align*}
    For any $D$ satisfying
    \begin{align*}
        D \leq \tilde{ \mathcal{O}} 
         \br{ (n+a)^{1-2^{-d}} m^{1+2^{-(d-1)}}  \log^{2}(1/\ve) },
    \end{align*}
    the class $\C$ satisfies condition (2) in \cref{lemma: Learning_Skeleton} with $d\gets D$ and $\ve_2\gets\ve$.
    That is, for any Choi representation $\Choi{\Phi} \in \C$, we have
    \begin{align}\label{eq:norm2-approximation-1}
        \min_{M \in \Matrix^{\leq D}} \norm{M-\Choi{\Phi}}_2 \leq \ve.
    \end{align}
    In fact, we have the stronger conclusion that \cref{eq:norm2-approximation-1} holds within the spectral distance.
    That is, in the approximate degree parlance, for any $\Choi{\Phi} \in \C$, we have
    \begin{align*}
         \degeps{\ve}{\Choi{\Phi}} \leq 
         \tilde{ \mathcal{O}} 
         \br{  (n+a)^{1-2^{-d}} m^{1+2^{-(d-1)}}  \log^{2}(1/\ve) } .
    \end{align*}
\end{prop}

To prove \cref{thm: QLC0-approximation-2}, this work requires some results in \cite{ADOY24}.
We therefore reformulate the results accordingly.

\begin{lemma} [Approximate degree for state, {\cite{Anshu2023concentrationbounds}, \cite[Corollary 3.2]{ADOY24}}]
    Let $\ket{\psi}$ be an $\ell$-qubit pure state. Then for any $r \in (\sqrt{n}, n)$ and $\ve = 2^{-\frac{r^2}{2^8 n}}$. It holds that
    \begin{align*}
        \degeps{\ve}{\ketbra{\psi}^{\otimes n}} \leq lr
    \end{align*}
\end{lemma}

We only care about the EPR state with the following corollary.

\begin{corollary}  \label{cor: EPR_approximate_lemma}
    For any integer $n>0$ and $0 < \ve < 2^{-1/2^{8}}$,
    \begin{align*}
        \degeps{\ve}{2^{-n}\ketbra{\EPR{n}}} \leq 
        \mathcal{O}\br{ \sqrt{n \cdot \log(1/\ve) } }
    \end{align*}
\end{corollary}

Note that when $\ve < 2^{-n}$, the inequality holds trivially since $\deg \br{2^{-n}\ketbra{\EPR{n}}} \leq 2n$.

\begin{lemma} [{\cite[Lemma 2.12]{ADOY24}}] \label{lemma: QAC0_approximate_lemma_ADOY}
   Given integers  $m \leq n$ and $ 0 < \ve < 1$, it holds that for $M \in \Matrix_{2^n}$ and any $2^m \times 2^m$ density operator $\varphi$, 
    \begin{align*}
        \degeps{\ve}{\partrace{n-m+1, \ldots, n}{(\id \otimes \varphi) M}} \leq  \degeps{\ve}{M}
    \end{align*}
\end{lemma}

\begin{proof}[Proof of \cref{thm: QLC0-approximation-2}]
With \cite[Fact 7.1]{ADOY24},
\begin{align*}
    2^{-m} \Choi{\Phi} = \bra{\psi} \br{\id \otimes U^T} \br{2^{-m}\ketbra{\EPR{m}} \otimes \id_{n+a-m}} \br{\id \otimes \bar{U}} \ket{\psi}.
\end{align*}

Let $\ve' = \ve/3$.
Apply \cref{cor: EPR_approximate_lemma} with parameter $\ve\gets\ve'$,
we get that there is an operator $M$ with degree $\ell = \mathcal{O}\br{\sqrt{m \log(1 / \ve)}}$ such that
\begin{align*}
    \norm{M - \br{2^{-m}\ketbra{\EPR{m}} \otimes \id} } \leq \ve'.
\end{align*}
Then, apply \cref{thm: QAC0-multi-layer} with parameters $r\gets\tilde{\mathcal{O}}(\log(1/\ve'))$ and $A\gets M$,
%Then, choose $\ve' = \ve / 3, r = \tilde{\mathcal{O}}(\log(1/\ve))$ and $A = M$
%in \cref{thm: QAC0-multi-layer},
where the condition $m \cdot \log(1/\ve) \leq n^{C_d}$ ensures that the condition of the \cref{thm: QAC0-multi-layer} holds.
There exists an operator $\tilde{M}$ with degree 
\begin{align*}
    \mathcal{O} \br{ (n+a)^{1-2^{-d}} m^{2^{-d}}  \log^{1+2^{-d}}(1/\ve)}
\end{align*}
such that
\begin{align*}
    \norm{\tilde{M} - \br{\id \otimes U^T} M \br{\id \otimes \bar{U} }} \leq \ve'.
\end{align*}

Now,
\begin{align*}
    &\;\; \norm{\tilde{M} - \br{\id \otimes U^T} \br{2^{-m}\ketbra{\EPR{m}} \otimes \id} \br{\id \otimes \bar{U}}} \\
    &\leq \norm{\tilde{M} - \br{\id \otimes U^T} M \br{\id \otimes \bar{U}}} 
    + \norm{\br{\id \otimes U^T} \br{M - 2^{-m}\ketbra{\EPR{m}} } \br{\id \otimes \bar{U}}} \\
    & \leq \ve' + 2\ve' = \ve.
\end{align*}

As a conclusion,
\begin{align*}
    &\;\; \degeps{\ve}{\br{\id \otimes U^T} \br{2^{-m}\ketbra{\EPR{m}} \otimes \id_{n+a-m}} \br{\id \otimes \bar{U}}} \\
    &= \tilde{ \mathcal{O}} \br{ (n+a)^{1-2^{-d}} m^{2^{-d}}  \log^{1+2^{-d}}(1/\ve) }.
\end{align*}
With \cref{lemma: QAC0_approximate_lemma_ADOY}, 
\begin{align*}
    \degeps{\ve}{2^{-m}\Choi{\mathcal{E}_{m, U,\psi}}} \leq \degeps{\ve}{\br{\id \otimes U^T} \br{2^{-m}\ketbra{\EPR{m}} \otimes \id_{n+a-m}} \br{\id \otimes \bar{U}}}.
\end{align*}
Reorganizing it as
\begin{align*}
    \degeps{\ve}{\Choi{\mathcal{E}_{m, U,\psi}}} \leq \tilde{ \mathcal{O}} \br{ (n+a)^{1-2^{-d}} m^{1+2^{-(d-1)}}  \log^{2}(1/\ve) }.
\end{align*}
The proof completes here.
\end{proof}

We are now ready to prove \cref{thm: QLC0-approximation-2-true}.
\begin{proof} [Proof of \cref{thm: QLC0-approximation-2-true}]
Plugging \cref{thm: QLC0-approximation-2} and \cref{lemma: approx-channel-2-norm} into \cref{lemma: Learning_Skeleton}, we conclude that Algorithm \ref{alg: Channel-Tomography} is indeed an agnostic learning algorithm for \QLCz\ channels, thus proving \cref{thm: QLC0-approximation-2-true}. 
%The proof of \cref{thm: QLC0-approximation-2-true} completes.
\end{proof}

%\cref{thm: QLC0-approximation-2-true} exhibits significant dependence on $\ve$. Building upon the results of \cite{ADOY24}, we next derive an agnostic learning algorithm that is robust to $\ve$ but suboptimal in the degree parameter.

To prove \cref{thm: QAC0-agnostic-algorithm-two}, we use the following proposition in place of \cref{thm: QLC0-approximation-2}.
\begin{prop}\label{prop: QLC0_approximate_lemma_1}
    Given $0 < \ve < 1$, $n,m,d \geq 1$ and $a \geq 0$, and
    assume $\C$ is a set of Choi representations of $n \to m$ quantum channels implemented by depth-$d$ \QACz\ circuits with $a = \mathrm{poly}(n)$ ancilla initialized in a pure state $\psi$. 
    % and implement the quantum channel
    % \begin{align*}
    %      \mathcal{E}_{m, U,\psi}(\rho) = \partrace{[m]^c}{U(\rho \otimes \psi) U^{\dagger}}.
    % \end{align*}
    For any $D$ satisfying
    \begin{align*}
        D \leq \tilde{ \mathcal{O}} 
         \br{ (n+a)^{1-3^{-d}} m^{3^{-d} + 1/2}  \log(1/\ve) },
    \end{align*}
    the class $\C$ satisfies condition (2) in \cref{lemma: Learning_Skeleton} with $d\gets D$ and $\ve_2\gets\ve$.
    I.e., for any $\Choi{\Phi} \in \C$, we have
    \begin{align}
        \min_{M \in \Matrix^{\leq D}} \norm{M- \Choi{\Phi}}_2 \leq \ve.
    \end{align}
\end{prop}

The proof of this proposition is highly similar to that of \cref{thm: QLC0-approximation-2}, and the argument is completed by simply replacing \cref{thm: QAC0-multi-layer} with \cite[Lemma 3.5]{ADOY24}.

% We have demonstrated how to use previous results to give a learning algorithm for \QACz. By replacing the lemmas therein with their counterparts presented in this work, we can extend the learning to \QACz\ circuits with more ancilla. However, due to the restriction on 
% $r$ in Theorem \ref{thm: QAC0-multi-layer}, the resulting learning algorithm will have a constraint on the number of outputs.

%\subsection{Applications in Tolerant Testing} \label{subsec: tolerant-testing}
\section{Tolerant Testing for \texorpdfstring{\QLCz}{QLC0} channels}\label{sec:tolerant-testing}
% \paragraph{Diamond Norm}

% We have used the spectral norm as the measure of approximation for circuits. However, in applications, we have used only the properties of the $2$-norm. A natural question is to ask whether we can get a learning result  under $p$-norm, or even the spectral norm?

% According to Lemma \ref{lemma: Learning_Skeleton}, the main difficulty lies in solving the low-degree approximation for general quantum channels under stronger norms. For $L_p$ norms, even in the classical setting, this problem currently has neither known algorithms nor established lower bounds. Several works \cite{Lp-regression-1}, \cite{Lp-regression-2} have considered the $L_p$-regression problem in the univariate case, while the multivariate scenario remains unexplored.

% Existing works such as \cite{Fawzi:2023oub} and \cite{PhysRevLett.132.180805} have demonstrated the hardness of learning Pauli channels. The work \cite{vasconcelos2024learning} shows that learning unitary  in \QACz\ under the diamond norm requires an exponential number of samples, even without any ancillary qubits. However, this result still does not establish the further hardness (for the few-output case) of learning low-degree channels or \QLCz\ channels. 
% In the following section we will demonstrate how to leverage these results to derive stronger conclusions.

Here, we exhibit an application of our agnostic learning algorithm in tolerant testing.

\begin{definition}[$\ve$-gap Tolerant testing for channels] \label{def: Tolerant-testing}
    Let $0 < \ve_1 < \ve_2 < 1$ such that $\ve_2 - \ve_1 \geq \ve$.
    Let $\Phi$ be an unknown channel, and $\C$ a set of Choi representations of quantum channels.
    Assuming that one of the following two cases is true, the task is to determine which one.
    \begin{enumerate}
        \item $\min_{C \in \C} \norm{C - \Choi{\Phi}}_2 \leq \ve_1$.
        \item $\min_{C \in \C} \norm{C - \Choi{\Phi}}_2 \geq \ve_2$.
    \end{enumerate}
\end{definition}

By augmenting the agnostic learning algorithm with an estimator for the distance between the target channel, we can directly obtain a tolerant testing algorithm.

\begin{prop}
    Suppose $\mathcal{A}$ is an $(\ve, \delta)$-agnostic learning algorithm for $\C$ under normalized Frobenius norm, then there is a $2\ve$-gap tolerant testing algorithm with the same sample and computational complexity.
\end{prop}

\begin{proof}
    Suppose $\mathcal{A}$ outputs $L$, then one just checks whether
    \begin{align}
        \norm{L - \Choi{\Phi}}_2 \geq \frac{\ve_1 + \ve_2}{2}
    \end{align}
    or not.
    For the yes case, determine that it falls under the case $\geq \ve_2$. Conversely, determine that it falls under the case $\leq \ve_1$. 
    The property of agnostic learning
    \begin{align}
        \norm{L - \Choi{\Phi}}_2 \leq \min_{C \in \C} \norm{\Choi{C} - \Choi{\Phi}}_2 + \ve
    \end{align}
    ensures correctness.
\end{proof}

By slightly decreasing the value of $\ve$, the algorithm can further accommodate errors in distance computation.
In particular, by using low-degree operators as a bridge, we can efficiently compute the distance, leading to the following corollary:

\begin{corollary}\label{cor:tolerant-testing}
    Assume $\C$ is the Choi representation of a family of \QLCz\ $n \to \operatorname{polylog}(n)$ channels. 
    There exists an  $1/\operatorname{poly}(n)$-gap tolerant testing algorithm
    with sub-exponential sample and time complexity. 
\end{corollary}

\begin{proof}
    By applying the agnostic learning algorithm in Algorithm \ref{alg: Channel-Tomography}, we obtain an $L$.
    Recall that $\norm{L - \Choi{\Phi}^{\leq d}}_2 \leq \ve$, 
    after estimating $\norm{\Choi{\Phi}}_2$ and obtaining a value $v$ such that $\abs{ v - \norm{\Choi{\Phi}}_2 } \leq \ve$,
    we have
    \begin{align*}
        \norm{L - \Choi{\Phi}}_2 
        &\leq \norm{L - \Choi{\Phi}^{\leq d}}_2 + \norm{\Choi{\Phi} - \Choi{\Phi}^{\leq d}}_2 \\
        &= \ve + \sqrt{\norm{\Choi{\Phi}}_2^2 - \norm{\Choi{\Phi}^{\leq d}}_2^2} \\
        &\leq \ve + \sqrt{(v+\ve)^2 - \br{\norm{L}_2 - \ve}^2} \\
        &\leq \sqrt{v^2 - \norm{L}_2^2} + 2\sqrt{\ve} + \ve \\
        &\leq \sqrt{v^2 - \norm{L}_2^2} + 3\sqrt{\ve}.
    \end{align*}

    And conversely, since $L$ is low-degree,
    \begin{align*}
        \norm{L - \Choi{\Phi}}_2 
        &\geq \norm{\Choi{\Phi} - \Choi{\Phi}^{\leq d}}_2 \\
        &=\sqrt{\norm{\Choi{\Phi}}_2^2 - \norm{\Choi{\Phi}^{\leq d}}_2^2} \\
        &\geq \sqrt{(v-\ve)^2 - \br{\norm{L}_2 + \ve}^2} \\
        &\geq \sqrt{v^2 - \norm{L}_2^2} - 2\sqrt{\ve}.
    \end{align*}
    Still, since $L$ is low-degree, $\norm{L}_2$ is computable.
    Thus, if we apply agnostic learning algorithm for $
    \ve' = o\br{(\ve_2- \ve_1)^2}$, we have accomplished the estimation of the distance with error $o \br{\ve_2 - \ve_1}$, thereby completing the tolerant testing .
\end{proof}

Note that if we can approximate the value of $\norm{\Choi{\Phi}^{\leq d}}_2$ more efficiently (without requiring knowledge of its approximate form), we could derive a more efficient testing algorithm.

% \subsection{Model: without trace}

% Naturally, we notice that when dealing with ancilla, it is difficult to provide better approximation results with spectral norm under partial trace. Therefore, we turn to consider the following model: we do not require taking partial trace of the output qubits, instead, all output qubits are given as results. That is,

% \begin{align}
%     \mathcal{E}_{U,\psi}(\rho) = U(\rho \otimes \psi) U^{\dagger}.
% \end{align}

\section{Hardness on Learning \texorpdfstring{\QACz}{QAC0} Channels}\label{sec:hardness}

In this section, we prove our hardness results of learning quantum channels in terms of the spectral norm.
\begin{restatable} [Hardness of \QACz\ channel learning within spectral norm] {theorem}{CIVmain}
\label{thm: hardness-on-QAC0-few}
    Given $n, m,a,d \geq 1$.
    Assume $\C$ is the set of all Choi representations of $n \to m$ channels implemented
    by a depth-$d$ \QACz\ unitaries $U$ with $a$ ancilla, where $a \geq m$.
    Then, given an unknown channel $\Choi{\Phi} \in \C$,
    learning $\Choi{\Phi}$ up to a spectral distance $\frac{1}{n}$
    with probability at least $1-\frac{1}{n^2}$ requires $\exp \br{\Omega(n)}$ queries.
\end{restatable}

% \begin{remark}
%    For technical reasons, we need the requirement that $a \ge m$.
% \end{remark}

This hardness is proved via a reduction to unitary hardness.
%an algorithm on how to learn unitaries generated by \QACz\ circuits with linear ancilla. This can be viewed as a complement to the previous sections, where the number of output qubits was $n$. 
%We initially believed that this algorithm was efficient.
%During our algorithmic analysis, we observed that while this approach still does not resolve the \QLCz\ unitary learning problem, it unexpectedly reveals 
In \cref{subsec:hardness-reduction}, we present the reduction, and the proof is given in \cref{subsec:hardness-results}.

\subsection{Hardness Reduction}\label{subsec:hardness-reduction}

We now introduce the key lemma to be used in the reduction, which enables separate processing of each output qubit and ultimately synthesizes the target unitary matrix.

\begin{definition} [Local inversion]
\label{def: local inversion}
    $V_i$ is a local inversion of $U$ for the $i$-th qubit  if and only if
    \begin{align*}
        UV_i = U_{-i} \otimes \id_i
    \end{align*}
    for some unitary $U_{-i}$ acting only on $\set{i}^c$.
\end{definition}

\begin{lemma} [{\cite[Eq. (5)]{HLB+24}}]
\label{lemma: local inversion}
    Given a unitary $U \in \Matrix_n$,
    for each $i\in[n]$, suppose $V_i$ is a local inversion of $U$ for the $i$-th qubit, then
    \begin{align}
        U \otimes U^{\dagger} = S \cdot \prod_{i=1}^n \br{V_i \cdot S_i \cdot V_i^{\dagger}}
    \end{align}
    where $S_i$ is the SWAP operator for $i$-th and $(i+n)$-th qubit, $S = \prod_{i=1}^n S_i$ and $V_i \cdot S_i \cdot V_i^{\dagger}$ are the abbreviations of $(V_i \otimes \id) \cdot S_i \cdot (V_i^{\dagger} \otimes \id)$.
\end{lemma}

The following lemma allows us to focus solely on estimating Heisenberg-evolved Pauli observables.

\begin{lemma}
\label{lemma: Heisenberg-evolved Pauli observables}
    For $x\in\set{0,1,2,3}$ and $i \in [n]$, let $\Base{x}^{(i)}$ be an $n$-qubit operator that applies the Pauli operator $\Base{x}$ on $i$-th qubit and $\id$ on other qubits. That is
    \begin{equation*}
        \Base{x}^{(i)} = \id^{\otimes i-1}\otimes\Base{x}\otimes\id^{\otimes n-i-1}.
    \end{equation*}
     Then,
    \begin{align}
        V_i S_i V_i^{\dagger} = \frac{1}{2} \sum_{x \in  \set{0,1,2,3}}  V_i \Base{x}^{(i)} V_i^{\dagger} \otimes \Base{x}^{(i+n)}
    \end{align}
    where $S_i$ is the SWAP operator for $i$-th and $(i+n)$-th qubits. 
\end{lemma}
\begin{remark}\label{rk:heisenberg}
    $V_i \Base{x}^{(i)} V_i^{\dagger}$ are called the Heisenberg-evolved Pauli observables. 
\end{remark}

\begin{proof}
    With the fact $S_i = \frac{1}{2} \sum_{x \in \set{0,1,2,3}} \br{\Base{x}^{(i)} \otimes \Base{x}^{(i+n)}}$, 
    \begin{align*}
        V_i S_i V_i^{\dagger} 
        =  \frac{1}{2} \sum_{x \in \set{0,1,2,3}} (V_i \otimes \id) \br{\Base{x}^{(i)}\otimes \Base{x}^{(i+n)}} (V_i^{\dagger} \otimes \id) 
        = \frac{1}{2} \sum_{x \in  \set{0,1,2,3}}  V_i \Base{x}^{(i)} V_i^{\dagger} \otimes \Base{x}^{(i+n)}.
    \end{align*}
\end{proof}
%\pnote{where is the proof or reference?} 

We then present the reduction directly in Reduction \ref{alg: Unitary-Tomography}. The key intuition is that to learn Heisenberg-evolved Pauli observables, we can attempt to learn the channel: $\rho \to V_i\br{\rho \otimes \id} V_i^{\dagger}$. Notably, the dual of this channel precisely corresponds to the sub-channel that restricts the circuit output to the $i$-th qubit.

\begingroup
\renewcommand{\algorithmcfname}{Reduction}

\begin{algorithm}[H]
\SetKwInOut{Input}{Input}
\SetKwInOut{Output}{Output}

\caption{\algname{Unitary-Reduction}($\Phi_U, \mathcal{A}$, $\delta$)}
\label{alg: Unitary-Tomography}
\Input{Given the Choi state of an $n$-input \QACz\ quantum channel $\Phi_U$,  and access to an algorithm $\mathcal{A}$ which outputs an approximation of a Choi representation when given access to the Choi state, and a real number $0<\delta < 1$.}	
\Output{An approximation $Q \approx U \otimes U^{\dagger}$ w.p. $1-\delta$.}

\begin{algorithmic}[1]
\item Repeat the following for $i \in \set{1,2,\cdots,n}$;
\vspace{-\topsep}
\begin{enumerate}[A]
\item 
Let $V_i$ be a local inversion of $V$ corresponding to the $i$-th qubit and $\Phi_{V_i}(\rho) = \partrace{-i}{V_i \rho V_i^{\dagger}}$ be the $n \to 1$ channel with only the $i$-th qubit reserved.
Given Choi state $\Chois{\Phi_U}$, trace out all but the $i$-th qubit to get $\Chois{\Phi_{V_i}}$.

\item Learn an approximation $M_i$ for $\Choi{\Phi_{V_i}}$ using algorithm $\mathcal{A}$ with access to $\rho(\Phi_{V_i})$ and failure probability $\delta/n$.

\item Learn $Q_{i,x}$ as an approximation of Heisenberg-evolved Pauli observables $V_i\Base{x}^{(i)}V_i^{\dagger}$ (defined in \cref{rk:heisenberg}): set $Q_{i,0} = \id$ and 
$Q_{i,x} =  \partrace{1}{M_i^T (\Base{x} \otimes \id)}$ for $x = 1,2,3$.

\item Sewing the local inversion $V_i S_i V_i^{\dagger}$ from Heisenberg-evolved Pauli observables: set $Q_i = \sum_{x} Q_{i,x} \otimes \Base{x}^{(i+n)}$.
\end{enumerate}
\vspace{-\topsep}
\item Return $Q = S \cdot \prod_{i=1}^n Q_i$.

\end{algorithmic}
\end{algorithm}

\endgroup

Unlike \cite{vasconcelos2024learning} and \cite{HLB+24}, we perform no operations during step B of our protocol.

%The algorithm enables us to learn the global unitary $U$ with the capability to learn local channels $\set{\Phi_{V_i}}$.
%\pnote{what are global unitary, local channels?} 
In other words, this reduction transfers unitary hardness to channel hardness.

%We explain in the next subsection how this algorithm serves as a reduction. 
%\pnote{what is this paragraph about?}Additionally, we have discussed how this reduction can be practically applied to shallow quantum circuit learning. Since this discussion is not directly relevant to the current section's content, we have placed it in the appendix.

\begin{theorem} \label{thm: hardness-unitary-to-channel}
    Given an integer $n>0$ and real numbers $ 0<\ve, \delta < 1$, let $\mathcal{U}$ be a set of $n$-qubit unitaries.
    Suppose given access to any channel $\Phi$ of the form $ \partrace{[n-1]}{\Phi_U}$ with single qubit output where $U \in \mathcal{U}$, algorithm $\mathcal{A}$ outputs a hypothesis $H$ w.p. at least $1-\delta$ such that
    \begin{align}
        \norm{H - \Choi{\Phi}} \leq \ve.
    \end{align}
    Then, there exists an algorithm $\mathcal{A'}$,  given access to a channel $\Phi_U$ with $U \in \mathcal{U}$, algorithm $\mathcal{A'}$ performs Reduction \ref{alg: Unitary-Tomography} using $\mathcal{A}$ as an oracle in step B, then outputs a hypothesis $H'$ w.p. at least $1- n\delta$ such that
    \begin{align}
        \norm{H' - U \otimes U^{\dagger}} \leq 9 \cdot n\ve.
    \end{align}
    Moreover, if $\mathcal{A}$ has sample complexity $N(\ve, \delta, n)$, then $\mathcal{A}'$ has sample complexity
    \begin{align*}
        N'(\ve,\delta,n) = n \cdot N(\ve / 9n, \delta / n, n).
    \end{align*}
\end{theorem}

The above theorem essentially demonstrates that if the unitaries of a circuit class are hard to learn, then their corresponding channels should also exhibit learning hardness. 

\begin{proof} [Proof of \cref{thm: hardness-unitary-to-channel}]

Let $V$ denote the implicitly declared unitary matrix, $V_i$ be the local inversion (\cref{def: local inversion}) of $V$ corresponding to the 
$i$-th qubit and $\Phi_{V_i}$ be the $n \to 1$ channel with only the $i$-th qubit reserved.

To begin with the analysis of the reduction, 
we assume that in step B, one can learn $Q_{i,x}$ such that 
\begin{align*}
    \norm{M_i - J(\Phi_{V_i})} \leq \ve.
\end{align*}
And starting from now on, we fix $\ve$ and define $\ve_V$ which will subsequently be derived from $\ve$.

    Now, assuming that step 1 of the reduction has been successfully completed,
by a hybrid argument, with a straightforward calculation we can show that
under the condition where step 1 provides a valid estimation, $Q$ is indeed a valid approximation of $U \otimes U^{\dagger}$.
Formally, suppose $\norm{Q_i -V_i S_i V_i^{\dagger}} \leq \ve_V < 1/n$, then 
\begin{equation}\label{eq:6994e8c9-f0ce-4b85-974b-0a3796ea6e25}
  \norm{Q - U\otimes U^{\dagger}} \leq 3 n \ve_V.
\end{equation}
We defer the proof to the \cref{sec:proofs-appendix}.

The remaining task is to justify the correctness of step 1, which is also the most intriguing part. 

We first prove that in step A, performing the partial trace on the Choi representation $\Choi{\Phi_U}$ indeed yields the Choi representation $\Choi{\Phi_{V_i}}$ corresponding to the sub-channel acting only on the $i$-th qubit. This is divided into two steps: the first step verifies the relationship between $\Phi_U$ and $\Phi_V$.  

\begin{align*}
    \Choi{\Phi_U} 
    &= (\id \otimes \bra{0_a}) \br{(\id \otimes U) \ketbra{\EPR{n}} (\id \otimes U^{\dagger})} (\id \otimes \ket{0_a}) \\
    &= (\id \otimes V) \ketbra{\EPR{n}} (\id \otimes V^{\dagger}) \\
    &= \Choi{\Phi_V}
\end{align*}
The second equality follows from the clean computation assumption.

The second step verifies that the partial trace of the Choi state indeed yields the Choi state corresponding to the sub-channel, since
\begin{align*}
    \partrace{-i}{\Choi{\Phi_V}} 
    &= \partrace{-i}{ (\id \otimes U' \otimes \id_i) (\id \otimes V_i)  \ketbra{\EPR{n}}  (\id \otimes V_i)^{\dagger}  (\id \otimes U' \otimes \id_i)^{\dagger}}\\
    &= \partrace{-i}{  (\id \otimes V_i)  \ketbra{\EPR{n}} (\id \otimes V_i)^{\dagger} } \\ 
    &= \Choi{\Phi_{V_i}}.
\end{align*}

From the hypothesis, in step B, we have
\begin{align*}
    \norm{M_i - \Choi{\Phi_{V_i}}} \leq \ve.
\end{align*}

Let us focus on the step C. We review some facts originating from quantum information theory (see the book of Watrous \cite{watrous2018theory} for more information): if an $n\to 1$ channel $\Phi$ has the form $\rho \to \partrace{-i}{V \rho V^{\dagger}}$, then its dual channel $\Phi^*$ will take the form $\rho_i \to V^{\dagger} (\rho_i \otimes \id_{-i}) V$. Moreover, the Choi representation of the dual channel and that of the original channel satisfy the following relation: 
\begin{align*}
    \Choi{\Phi}^T = \Choi{\Phi^*}
\end{align*}
Thus,
\begin{align*}
    \norm{M_i^T - \Choi{\Phi_{V_i}^*} } = \norm{M_i - \Choi{\Phi_{V_i}}} = \norm{M_i^T - \Choi{\Phi_{V_i}}^T } \leq \ve.
\end{align*}

That is, $M_i^T$ is actually the approximation to the Choi representation of $\rho_i \to V^{\dagger} (\rho_i \otimes \id_{-i}) V$. This channel is of importance, since when inputting $\Base{x}$, the output corresponds precisely to the desired 
Heisenberg-evolved Pauli observables. By the fact 
\begin{align}
    \Phi(X) = \partrace{\text{input}}{\Choi{\Phi}(X^T \otimes \id)},
\end{align}
we have:
\begin{align*}
    \norm{Q_{i,x} - V_i \Base{x}^{(i)} V_i^{\dagger}}
    &= \norm{\partrace{1}{M_i^T (\Base{x} \otimes \id)} - \partrace{1}{\Choi{\Phi_i^*} (\Base{x} \otimes \id)}}  \\
    &\leq  2\norm{M_i^T (\Base{x} \otimes \id) - \Choi{\Phi_i^*} (\Base{x} \otimes \id)} \\
    &\leq  2\norm{M_i^T - \Choi{\Phi_i^*}} \cdot \norm{\Base{x} \otimes \id} \\
    &\leq 2 \ve.
\end{align*}

Finally, for step D, with \cref{lemma: Heisenberg-evolved Pauli observables},
\begin{align*}
    \norm{Q_i - V_i S_i V_i^{\dagger}} &\leq \frac{1}{2} \sum \norm{Q_{i,x} \otimes \Base{x}^{(i+n)} - V_i \br{\Base{x}^{(i)} \otimes \Base{x}^{(i+n)}} V_i^{\dagger}} \\
    &\leq \frac{1}{2} \sum \norm{Q_{i,x}- V_i \Base{x}^{(i)} V_i^{\dagger}} \cdot \norm{\Base{x}^{(i+n)}} \\
    &\leq \frac{1}{2} \cdot 3 \cdot 2\ve = 3\ve
\end{align*}
The final constant is 3, since we can always output perfect replicas on the
$\id$ case. Ultimately, we obtain $\ve_V = 3\ve$ and the total error distance $9 \cdot n\ve$. 

This completes the proof of \cref{thm: hardness-unitary-to-channel}.
\end{proof}

\subsection{Hardness Results}\label{subsec:hardness-results}

Let us briefly recall the background of the reduction, where we take the same setting as in Vasconcelos and Huang's work~\cite{vasconcelos2024learning}.
Suppose $U$ is a unitary generated by a circuit with $(n+a)$ qubits.
We only consider circuits where the ancilla qubits are initialized to the $\ket{0_a}$ state and the computation is clean. Since the computation is clean, $U$ implicitly characterizes a unitary matrix $V$ acting on $n$ input qubits.
\begin{align}
    U \br{\ket{\varphi} \otimes \ket{0^a}} = \br{V\ket{\varphi}} \otimes \ket{0^a}.
\end{align}

While it may appear that learning a unitary with ancilla is equivalent to the no ancilla case, it is crucial to note that introducing ancilla changes the properties of the unitary (like degree). We may use \QACz\ unitary $U$ with $a$ ancilla to refer to the implicit unitary $V$.

Now we are ready to prove \cref{thm: hardness-on-QAC0-few}. %\pnote{Add reference here }
Let us first examine the learning hardness of \QACz\ unitary.
We use a result of Vasconcelos and Huang \cite{vasconcelos2024learning}, tailored to our needs.
For completeness, we provide a proof.
\begin{lemma} [Tailored from {\cite[Propsition 7]{vasconcelos2024learning}}] \label{thm: QAC0 is hard to learn}
    Consider an unknown $n$-qubit unitary $U$ generated by a depth-$d$ \QACz\ 
    circuit with $a$ ancilla.
    Then, 
    distinguishing whether $U \otimes U^{\dagger}$ equals the $\id$ or is $\frac{1}{3}$-far from $\id$ in diamond distance with  probability $2/3$ requires $\exp \br{\Omega(n)}$ queries.
\end{lemma}
\begin{proof}
    For $x, y \in \set{0,1}^n$, let $U_x$ be the unitary,
    \begin{align}
        U_x \ket{y} = (-1)^{\delta_{xy} + 1} \ket{y}. 
    \end{align}
    which can be constructed as
    \begin{align*}
        U_x = \Base{\bar{x}} \CZG  \Base{\bar{x}}.
    \end{align*}

    $U_x$ is in the depth-$d$ \QACz\ circuit with ancilla even for $a = 0$ and $d = 1$. However, distinguishing $\id$ from one of $U_x$ has a Grover Search lower bound from Bennett, Bernstein, Brassard and Vazirani's work \cite{doi:10.1137/S0097539796300933} with queries at least $\exp \br{\Omega(n)}$. 
    This also means distinguishing $\id$ from one of $U_x \otimes U_x^{\dagger}$ has an exponential bound.
\end{proof}

Combining this hardness assumption for \QACz\ unitary with \cref{thm: hardness-unitary-to-channel}, we can establish hardness results even for channels with minimal input requirements.

\begin{lemma} [Hardness of \QACz\ $n \to 1$ channel learning with spectral distance] \label{prop: hardness-QAC0-one-output}
    Given integers $a \geq 0$ and $n,d \geq 1$.
    Assuming $\C$ is the set of Choi representations of $n \to 1$ channels induced by depth-$d$ \QACz\ unitaries $U$ with $a$ ancilla. Then, given an unknown Choi $\Choi{\Phi} \in \C$, learning $\Choi{\Phi}$ up to a spectral distance $\frac{1}{n}$ with probability at least $1-\frac{1}{n^2}$ requires $\exp \br{\Omega(n)}$ queries.
\end{lemma}

\begin{proof}
One minor technical detail to note is that after performing the reduction in \cref{thm: hardness-unitary-to-channel}, we obtain a hypothesis $H$ such that
\begin{align*}
    \norm{H - U \otimes U^{\dagger}} \leq 1/12.  
\end{align*}
To relate this result to the diamond norm, we locally compute the nearest unitary matrix $\tilde{U}$ of $H$ in the spectral norm (though this may require exponential computation time, it does not affect sample complexity).
At this point, we have 
\begin{align*}
    \norm{U \otimes U^{\dagger} - \tilde{U}} \leq \norm{U \otimes U^{\dagger} - H} + \norm{H - \tilde{U}} \leq 1/6.
\end{align*}
Since the small spectral norm distance between unitary matrices leads to a small  diamond norm distance (\cref{prop: diamond-is-less-than-spectral}), this completes the proof with
\begin{align*}
    \norm{\Phi_{U \otimes U^{\dagger}} - \Phi_{\tilde{U}}}_{\diamond} \leq 2\norm{U \otimes U^{\dagger} - \tilde{U}} \leq 1/3.
\end{align*}
\end{proof}

We now begin the proof of \cref{thm: hardness-on-QAC0-few}.
\begin{proof}[Proof of \cref{thm: hardness-on-QAC0-few}]

Suppose $\mathcal{S}$ is the set of channels induced by  depth-$1$ \QACz\ circuits with $n$ inputs, sharing no ancilla and taking the first qubit as output. 
From \cref{prop: hardness-QAC0-one-output}, we know that learning $\mathcal{S}$ requires exponential samples.

Fix a channel $\Phi \in \mathcal{S}$, let $\tilde{\Phi}(\rho) = \Phi(\rho) \otimes \ketbra{0_{m-1}}$. Since $a \geq m$, $\tilde{\Phi}$ can be implemented by a depth-$d$ \QACz\ circuit with $n$ inputs and $a$ ancilla.

Since
\begin{align*}
    \Choi{\tilde{\Phi}} = \Choi{\Phi} \otimes \ketbra{0_{m-1}},
\end{align*}
we now have
\begin{align*}
    \norm{(\id \otimes \bra{0_{m-1}}) M (\id \otimes \ket{0_{m-1}}) - \Choi{\Phi}} \leq \norm{M - \Choi{\tilde{\Phi}} }
\end{align*}

Thus, if we are able to have $\norm{M - \Choi{\tilde{\Phi}} } \leq 1/n$, by taking $\tilde{M} = (\id \otimes \bra{0_{m-1}}) M (\id \otimes \ket{0_{m-1}})$, we have $\norm{\tilde{M} - \Choi{\Phi}} \leq 1/n$.  Learning $\mathcal{S}$ needs exponential queries, therefore, learning \QACz\ channels given in the condition also  needs exponential queries.

This completes the proof of \cref{thm: hardness-on-QAC0-few}.
\end{proof}

The spectral norm of the Choi representations is almost $2^m$. 
However, our hardness result only demonstrates that approximation is hard within a distance of $1/n$. 
When the number of outputs is sufficiently large (like $\sqrt{n}$), compared to $2^m$,
the value $1/n$ is too small to be an appropriate error.
This inspires us to propose the other hardness result in the diamond norm, which is powerful with a large number of output qubits.

%\dnote{$2^n-1/3.$ TODO}

\begin{theorem} [Hardness of \QACz\ channel learning within diamond norm distance]
\label{thm: hardness-QAC0-diamond-norm}
    Given integers $n,m \geq 1$ and $a,d \geq 0$.
    Assuming $\mathscr{L}$ is the set of $n \to m$ channels induced by depth-$d$ \QACz\ circuits with $a$ ancilla. Then, given an unknown channel $\Phi \in \mathscr{L}$, learning $\Phi$ up to diamond norm distance $1/3$  with probability $2/3$ requires $\exp \br{\Omega(m)}$ queries.
\end{theorem}

\begin{proof}
    Notice that the lower bound in \cref{thm: QAC0 is hard to learn} relies on a class of hard unitary $\set{U_n}$ that can be implemented by \QACz\ circuits. We can directly embed a size $k$ unitary $U_k$ into a channel with $k$ outputs, thereby deriving corresponding lower bounds for \QACz\ channels within diamond norm distance.
\end{proof}

\subsection{Hardness of Finding the Nearest Low-Degree Operator}

In this subsection, we present the hardness results further reduced from the aforementioned hardness of \QACz\ channel learning.

We recall the following proposition: \QLCz\ circuits all admit low-degree approximations. Consequently, for a given quantum channel, finding its closest low-degree operator remains computationally hard.

\begin{theorem} [Hardness of finding the nearest low-degree operator]
\label{thm: hardness-find-low-degree-approximation}
    Assume $M \in \Matrix_{2^n}$.
    For degree $D = \Omega \br{\sqrt{n} \cdot \operatorname{poly}(\log n)}$,
    finding
    \begin{align}
         \operatorname{argmin}_{X \in \Matrix_{2^n}^{\leq D}} \norm{X - M}
    \end{align}
    up to a spectral distance $\frac{1}{n}$ with probability at least $1-\frac{1}{n^2}$ requires $\exp \br{\Omega(n)}$ queries.
\end{theorem}

%The remaining step is to reduce the $m$-qubit output case to the single-qubit output case.

\begin{proof}
    We focus on the case where the \QACz\ channel $\Phi_U$ has at most depth $1$, has zero ancilla and only outputs the first qubit.  
    With Proposition \ref{thm: QLC0-approximation-2},
    \begin{align*}
        \degeps{1/n}{\Choi{\Phi_U}} = \sqrt{n} \cdot \textrm{poly} (\log n),
    \end{align*}
    i.e.,
    \begin{align*}
        \exists X \in \Matrix_{2^n}^{\leq D}, \norm{X - \Choi{\Phi_U}} \leq \frac{1}{n}.
    \end{align*}

    Assuming the hypothesis of this theorem holds, the closest low-degree operator to $\Choi{\Phi_U}$ would yield a $1/n$-approximation to $\Choi{\Phi_U}$, thereby contradicting Lemma \ref{prop: hardness-QAC0-one-output}.
\end{proof}

% \subsection{Agnostic learning}

% Alternatively, we can characterize both the local-global property of the unitary matrices generated by circuits.

\bibliographystyle{alpha}
\bibliography{references}

\appendix
\section{Proofs}\label{sec:proofs-appendix}

\begin{proof}[Proof of \cref{fact:dilation-is-unitary}]

Let  $D_A = \sqrt{I - A^\dagger A}$ and $U=A^{\uparrow}$, then
\begin{align*}
    U = \begin{bmatrix} A & D_{A^\dagger} \\ D_{A} & -A^\dagger \end{bmatrix}.
\end{align*}

With a direct calculation,
\begin{align*}
 U^\dagger U 
= \begin{bmatrix}
A^\dagger A + D_A D_A & A^\dagger D_{A^\dagger} - D_A A^\dagger \\
D_{A^\dagger} A - A D_A & D_{A^\dagger} D_{A^\dagger} + A A^\dagger
\end{bmatrix}
= \begin{bmatrix}
\id & A^\dagger D_{A^\dagger} - D_A A^\dagger \\
D_{A^\dagger} A - A D_A & \id
\end{bmatrix}.
\end{align*}

This shows that we only need to prove $A^\dagger D_{A^\dagger} - D_A A^\dagger = D_{A^\dagger} A - A D_A = 0$.  
By symmetry, it suffices to prove $D_{A^\dagger} A - A D_A = 0$.
Define $B = D_A^2$ and $C = D_{A^\dagger}^2$.
Now, we use an inductive argument to prove that for all $n \ge 1$, we have
\begin{align*}
    AB^n = C^nA.
\end{align*}
For the base case where $n=1$, it is to check that $AB=CA = A - AA^\dagger A$.
Now for any $n \geq 2$, suppose $AB^{n-1} = C^{n-1}A$ and $AB=CA$.
Then,
\begin{align*}
    AB^n = (AB^{n-1})B = (C^{n-1}A)B= C^{n-1}(AB) = C^{n-1}CA = C^nA.
\end{align*}

Thus, by a limit argument, for all elementary  function $f: \mathbb{R}\to\mathbb{R}$, we have
\begin{align*}
    Af(B)=f(C)A.
\end{align*}
In particular, for $f = \sqrt{x}$, we conclude that $D_{A^\dagger} A - A D_A = 0$.

\end{proof}

\begin{proof} [Proof of \cref{lemma: Qac0-one-layer}]
    The proof is similar to the one in \cite{ADOY24}. 

    When $r > n/\ell$, we have $n^{1/2}\ell^{1/2}r^{1/2} \geq n$.
    So choosing $M = UAU^{\dagger}$ directly completes the proof.
    In the remaining proof, we can assume $r \le \sqrt{n / \ell}$.
    
    Recall that $\CZGr_{S_i}$ acts on $n_i = \abs{S_i}$ qubits on the set $S_i\in[n]$. 
    We  divide the \CZGate s into three parts with parameter $t = \sqrt{n/\ell} > r$.
    Let $T_0 = \set{i : n_i \leq t}, T_1 = \set{i : t < n_i \leq t^2}, T_2 = \set{i : t^2 < n_i}$ and $T = T_1 \cup T_2$. 
    Define $U_{T_0} = \bigotimes_{i \in T_0} \CZGr_{S_i}$ and $U_{T_1}, U_{T_2}, U_{T}$ similarly.
    Without loss of generality, assume $T = [k]$.

    The key idea is as follows:
    \begin{itemize}
        \item For gates in $T_0$, we directly use a lightcone argument.
        \item For gates in $T_1$, we perform a low-degree approximation and use the lightcone argument.
        \item For gates in $T_2$, we perform a low-degree approximation and use a degree argument.
    \end{itemize}
    
    For each $i$, let $\widetilde{\CZGr}_{S_i}$ be the low-degree approximation of $\CZGr_{S_i}$
    derived from \cref{corollary: CZ-unitary-approximate-lemma}.
    Let $\tilde{U}_{T_0} = \bigotimes_{i \in T_0} \CZGr_{S_i}^{\uparrow}, \tilde{U}_T = \bigotimes_{i \in T} \widetilde{\CZGr}_{S_i}^{\uparrow}$ and $\tilde{U} = \tilde{U}_S \otimes \tilde{U}_T$. 
    Also, let $A^{\Uparrow}$ be the operator dilation of $A$ with respect to the ensemble $\mathcal{S} = \set{S_1, \cdots, S_m}$.
    %The $i$-th operator dilation corresponds to the unitary dilation of $\tilde{\CZGr}_{S_i}$. 
    We prove
    \begin{align} \label{align: Qac0-one-layer-target-1}
        \norm{U^{\uparrow}A^{\Uparrow} (U^{\dagger})^{\uparrow} - \tilde{U} A^{\Uparrow} \tilde{U}^{\dagger}} \leq Cn \cdot 2^{-2^{-9}r}.
    \end{align}
    Moreover, we show that the top-left part of $U^{\uparrow}A^{\Uparrow} (U^{\dagger})^{\uparrow}$ is exactly $UAU^{\dagger}$,
    and taking $M$ to be the top-left part of $\tilde{U} A^{\Uparrow} \tilde{U}^{\dagger}$ satisfies the requirements \cref{eq:one-layer-requirement-1} and \cref{eq:one-layer-requirement-2} and will complete the proof.

    To begin with, we prove \cref{align: Qac0-one-layer-target-1} with a hybrid argument. 
    Let $U_{(i)} = \tilde{U}_S \otimes \bigotimes_{j=1}^i \widetilde{\CZGr}_{S_j}^{\uparrow} \otimes \bigotimes_{j=i+1}^k \CZGr_{S_j}^{\uparrow}$.
    Now, $U_{(0)} = U^{\uparrow}$ and $U_{(k)} = \tilde{U}$.
    For $1 \leq i \leq k$,
    \begin{align}
        \norm{U_{(i)} - U_{(i - 1)}} 
        &= \norm{\tilde{U_S}  \otimes \bigotimes_{j=1}^{i-1} \widetilde{\CZGr}_{S_j}^{\uparrow} \otimes \bigotimes_{j=i+1}^k \CZGr_{S_j}^{\uparrow} \otimes \br{\CZGr_{S_i}^{\uparrow} - \widetilde{\CZGr}_{S_i}^{\uparrow}}} \\
        &= \norm{ \CZGr_{S_i}^{\uparrow} - \widetilde{\CZGr}_{S_i}^{\uparrow} } \\
        &\leq C_1 2^{-2^{-9}r}
    \end{align}
    where $C_1$ is a constant and the last inequality holds by \cref{corollary: CZ-unitary-approximate-lemma}.

    Thus,
    \begin{align}
        \norm{U_{(0)} - U_{(k)}} &\leq \sum_{i=1}^{k} \norm{U_{(i)} - U_{(i - 1)}} \leq C_1n 2^{-2^{-9}r}.
    \end{align}

    Hence, choosing $C = 2C_1$,
    \begin{align}
        &\;\; \norm{U^{\uparrow}A^{\Uparrow} (U^{\dagger})^{\uparrow} - \tilde{U} A^{\Uparrow} \tilde{U}^{\dagger}} \\
        &\leq 2 \cdot \norm{U^{\uparrow} - \tilde{U}} \cdot  \norm{A^{\Uparrow}} \\
        &\leq 2 \cdot \norm{U^{\uparrow} - \tilde{U}} \cdot  \norm{A} \\
        &\leq Cn2^{-2^{-9}r} \cdot  \norm{A}.
    \end{align}
    where the second inequality uses Lemma \ref{lemma: spectral-norm-for-operator-dilation}.

    For Pauli matrix $\Base{\sigma}$ and some unitary dilation $\set{M_i^{\uparrow}}$ with sets $\set{S_i}$,
    if $\sigma_{S_i} = 0_{S_i}$, then $M_i^{\uparrow}$ acts on the identity matrix. So it gets canceled out with itself:
    \begin{align}
        \br{M_i^{\uparrow} \otimes \id} \Base{\sigma}^{\Uparrow} \br{\br{M_i^{\dagger}}^{\uparrow} \otimes \id} 
        = \Base{\sigma}^{\Uparrow}.
    \end{align}
    If $\sigma_{S_i} \neq 0_{S_i}$,
    \begin{align}
        &\;\; \br{M_i^{\uparrow} \otimes \id} \Base{\sigma}^{\Uparrow} \br{\br{M_i^{\dagger}}^{\uparrow} \otimes \id} \\
        &= \Base{S_i^c}^{\Uparrow} \otimes \br{
            \begin{bmatrix} M_i & ... \\ ... &...\end{bmatrix} \cdot
            \begin{bmatrix} \Base{S_i} & 0 \\ 0 & 0\end{bmatrix} \cdot
            \begin{bmatrix} M_i^{\dagger} & ... \\ ... &...\end{bmatrix}
        } \\
        &= \Base{S_i^c}^{\Uparrow} \otimes 
            \begin{bmatrix} M_i\Base{S_i} M_i^{\dagger}& ... \\ ... &...\end{bmatrix} 
    \end{align}

   % Here, since $M_i \cdot \Base{S_i} \cdot M_i^{\dagger}$ may introduce identity operators, we cannot guarantee that matrices with $ M_i \cdot \Base{S_i} \cdot M_i^{\dagger}$ solely in the top-left corner can be expressed as a dilation.  We need to directly extract the top-left part.
    
    If $\set{M_i}$ are unitary matrices, for both cases, it holds that $\Base{\sigma}^{\Uparrow} = \br{M_i\Base{\sigma}M_i^{\dagger}}^{\Uparrow}.$
    Hence, 
    \begin{align}
        U^{\uparrow}A^{\Uparrow} (U^{\dagger})^{\uparrow} = \br{UAU^{\dagger}}^{\Uparrow}.
    \end{align}

    As a result, $UAU^{\dagger}$ is the top-left part of $U^{\uparrow}A^{\Uparrow}(U^{\dagger})^{\uparrow}$.
    Now, we choose $M$ as the top-left part of $\tilde{U} A^{\Uparrow} \tilde{U}^{\dagger}$. 
    
    Fix a Pauli matrix $\Base{\sigma}$ with $|\supp{\sigma}| = d \leq \ell$ in $A$.
    When $\Base{\sigma}$ acts with $U_S$, it becomes $U_{S}^{\dagger} \Base{\sigma} U_S$. 
    Since $\CZGr_{S_i} \Base{\sigma} \CZGr_{S_i}^{\dagger} = \id$ if $S_i \cap \supp{\sigma} = \emptyset$,  $U_{S}^{\dagger} \Base{\sigma} U_S$ has non-trivial part with degree at most $d n_i \leq dt$. 

    When $\Base{\sigma}$ acts with $U_{T_1}$, similar to $U_S$, the sets disjoint with $\supp{\sigma}$ degrade to $\id$ in the sense of dilation. 
    Now, we just consider $\widetilde{\CZGr}_{S_i} \Base{\sigma} \widetilde{\CZGr}_{S_i}^{\dagger}$, it has degree at most $2d\sqrt{n_i r} \leq 2dt \sqrt{r}$ since one qubit can get at most $\sqrt{n_i r}$ degree respectively from left side and right side.

    When $\Base{\sigma}$ acts with $U_{T_2}$, we just add the $\sqrt{n_i r}$ to the degree.
    Recall that $\sum_i n_i \leq n$, from here, one can get degree at most
    \begin{align}
        \sum \sqrt{n_i r} = \sum n_i \sqrt{\frac{r}{n_i}}
        \leq n \sqrt{\frac{r}{t^2}} = \frac{n}{t} \sqrt{r}.
    \end{align}

    Combine the three cases together, the degree of $M$ is at most
    \begin{align}
        2 \ell  t\sqrt{r} + 2 \cdot \frac{n}{t} \sqrt{r} 
        = 2\sqrt{r} \br{\ell t + n/t}
        \leq 4n^{1/2}\ell^{1/2}r^{1/2}.
    \end{align}
    The last inequality holds due to $t = \sqrt{n/\ell}$.

    The spectral norm of the top-left part is bound by the spectral norm of the whole matrix, thus,
    \begin{align}
        \norm{UAU^{\dagger} - M} \leq Cn \cdot 2^{-2^{-9r}}.
    \end{align}
\end{proof}

\begin{proof}[Proof of \cref{eq:6994e8c9-f0ce-4b85-974b-0a3796ea6e25}]
    Observe that the condition $\norm{Q_i -V_i S_i V_i^{\dagger}} \leq \ve_V < 1/n$ contains that $\norm{Q_n} \leq 1 + \ve_V$.
    Let $Q_{\leq i} = \prod_{j=1}^i Q_j, V_{\leq i} = \prod_{j=i}^i V_i \cdot S_i \cdot V_i^{\dagger }$.
    With a straightforward calculation,
    \begin{align*}
         &\;\; \norm{Q - U\otimes U^{\dagger}} \\
         &=  \norm{S \cdot \prod_{i=1}^n Q_i - S \cdot \prod_{i=1}^n V_i \cdot S_i \cdot V_i^{\dagger }} \\
         &= \norm{\prod_{i=1}^n Q_i - \prod_{i=1}^n V_i \cdot S_i \cdot V_i^{\dagger} } \\
         &= \norm{Q_{<n}Q_n - V_{<n}Q_n + V_{<n}Q_n - V_{<n} \br{V_n  S_n V_n^{\dagger}} } \\
         &\leq \norm{Q_n} \cdot \norm{ Q_{<n} - V_{<n}}
         +  \norm{Q_n - \br{V_n  S_n V_n^{\dagger}}} \cdot \norm{V_{<n}} \\
        & \leq (1+\ve_V) \cdot \norm{ Q_{<n} - V_{<n}} + \ve_V \\
        &\leq (1+\ve_V)^n - 1 \leq 3n\ve_V
    \end{align*}
    where the first equality comes from \cref{lemma: local inversion} and the last inequality comes from the fact that $(1+x)^n - 1 \leq 3nx$ for $x \in (0,1/n)$.
\end{proof}

\end{document}